\documentclass[]{aa} 

\usepackage{natbib}
\bibpunct{(}{)}{;}{a}{}{,} % to follow the A&A style
\usepackage{epsfig}

\newcommand{\beq}{\begin{equation}} 
\newcommand{\eeq}{\end{equation}} 
\newcommand{\beqa}{\begin{eqnarray}} 
\newcommand{\eeqa}{\end{eqnarray}} 
\newcommand{\bea}{\begin{array}} 
\newcommand{\ea}{\end{array}} 

\newcommand{\dd}{{\rm d}}

\newcommand{\lag}{\langle} 
\newcommand{\rag}{\rangle} 

\newcommand{\ii}{{\rm i}}

\newcommand{\rhob}{\overline{\rho}}

\newcommand{\vk}{{\bf k}}

\newcommand{\vq}{{\bf q}}

\newcommand{\vx}{{\bf x}}

\newcommand{\tdelta}{{\tilde{\delta}}}

\newcommand{\tk}{{\tilde{k}}}
\newcommand{\tkappa}{{\tilde{\kappa}}}
\newcommand{\tmu}{{\tilde{\mu}}}

\newcommand{\tu}{{\tilde{u}}}
\newcommand{\tW}{{\tilde{W}}}

\newcommand{\vtheta}{\vec{\theta}}
\newcommand{\vell}{\vec{\ell}}

\newcommand{\deltaLc}{{\delta_{L*}}}

\newcommand{\cD}{{\cal{D}}}

\newcommand{\Pkappa}{P_{\kappa}}
\newcommand{\Bkappa}{B_{\kappa}}

\begin{document} 

\topmargin =0.1cm

\title{Modeling of weak-lensing statistics. I. Power spectrum and bispectrum.}    
\author{
Patrick Valageas\inst{1}
\and
Masanori Sato\inst{2}
\and
Takahiro Nishimichi\inst{3}
}   
\institute{
Institut de Physique Th\'eorique, CEA Saclay, 91191 Gif-sur-Yvette, France 
\and
Department of Physics, Nagoya University, Nagoya 464-8602, Japan
\and
Institute for the Physics and Mathematics of the Universe, University of
Tokyo, Kashiwa, Chiba 277-8568, Japan
}
\date{Received / Accepted } 
 
\abstract
{}
{We investigate the performance of an analytic model of the 3D matter distribution,
which combines perturbation theory with halo models, for weak-lensing
statistics.}
{We compare our predictions for the weak-lensing convergence
power spectrum and bispectrum with numerical simulations and fitting formulas
proposed in previous works.}
{We find that this model provides better agreement with simulations than
published fitting formulas. This shows that building on systematic and physically
motivated models is a promising approach. Moreover, this makes explicit the
link between the weak-lensing statistics and the underlying properties of the
3D matter distribution, as a function of scale $\ell$. Thus, we obtain the contributions
to the lensing power spectrum and bispectrum that arise from perturbative 
terms (complete up to one-loop) and nonperturbative terms (e.g., ``1-halo'' term).
Finally, we show that this approach recovers the dependence on cosmology
(for realistic scenarios).}
{}

\keywords{weak gravitational lensing; cosmology: theory -- large-scale structure of Universe}

\maketitle

\section{Introduction} 
\label{Introduction}

Weak gravitational lensing of background galaxies by foreground large-scale
structures, the so-called cosmic shear, is one of the best tools to probe the
nature of the main components of the Universe, such as
dark matter and dark energy.
Thus, weak lensing has the highest potential to constrain the properties of
dark energy among other cosmological observations, if the systematic
errors are well kept under control~\citep{Albrecht2006}.
To address questions about the nature of dark energy and the properties
of gravity on cosmological scales, various surveys are planned, 
such as the Hyper Suprime-Cam Weak Lensing
Survey~\citep{Miyazaki2006}\footnote{http://www.naoj.org/Projects/HSC/index.html},
the Dark Energy Survey~(DES)\footnote{http://www.darkenergysurvey.org/},
the Large Synoptic Survey Telescope~(LSST)\footnote{http://www.lsst.org/},  
Euclid~\citep{Refregier2010}\footnote{http://www.euclid-ec.org/},
and the Wide-Field Infrared Survey
Telescope~(WFIRST)\footnote{http://wfirst.gsfc.nasa.gov/}.

To exploit the full potential of future weak-lensing surveys, it will
be important to analyze data with adequate statistical measures and
tools. Particularly, one needs to properly take into
account the correlations of the observables between different angular scales and
redshifts, i.e., their
covariance matrices~\citep{Cooray2001b,Takada2009,Sato2009,Sato2011a}.
Furthermore, one has to use an appropriate likelihood function with given marginal
distributions~\citep{Sato2010,Sato2011}.

Since most of the useful cosmological information contained in the
cosmic shear signal is associated with small angular scales that are
affected by nonlinear clustering, 
we also need to include these nonlinear effects to accurately model the weak-lensing
statistics~\citep{Takada2004,Sato2009}.
Most researchers use fitting formulas based on numerical simulations or
phenomenological approaches but it would be useful to obtain analytical methods
that are more directly related to the cosmological parameters and
primordial fluctuations.

In this paper, we examine the performance of the theoretical
modeling of the 3D matter density distribution proposed by
\citet{Valageas2011d,Valageas2011e}, which is based on a
combination of perturbation theories and halo models.
We focus on the convergence power spectrum and
bispectrum, which are basic statistical measurements in weak
lensing studies~\citep[see,][for a recent method of lensing power spectrum
measurement]{Hikage2011}.
As compared with simple fitting formulas or direct numerical simulations, 
a significant advantage of our approach is that we can evaluate and
compare different contributions that can be measured in weak-lensing
surveys.
Since different contributions suffer from different theoretical
uncertainties, this is useful to estimate the accuracy that can be aimed
at in weak-lensing statistics, as a function of scales.
Furthermore, we find that our model provides better agreement with numerical
simulations than other existent models.

This paper is organized as follows.
In Sect.~\ref{Analytic} we first present our model for the 3D matter density
power spectrum and bispectrum. Next, we recall how this yields the weak
lensing convergence power spectrum and bispectrum through the Born
approximation.
We describe our numerical simulations and the data analysis
in Sect.~\ref{Numerical}. Then, we present detailed comparisons between the
simulation results, previous models, and our
theoretical predictions, for the convergence power spectrum in
Sect.~\ref{Convergence-power-spectrum}, and for the convergence bispectrum
in Sect.~\ref{Convergence-bispectrum}, considering the cases of both equilateral
and more general isosceles configurations.
We study the relative importance of the different contributions to the power spectrum
and bispectrum in Sect.~\ref{contributions}, arising from ``1-halo'', ``2-halo'', 
or ``3-halo'' terms.
Finally, we check the robustness of our model as we vary the cosmological 
parameters in Sect.~\ref{Cosmology} and we conclude in 
Sect.~\ref{Conclusion}.

\section{Analytic models}
\label{Analytic}

\subsection{3D matter power spectrum and bispectrum}
\label{3D-power}

Our Fourier-space normalizations for the density contrast, its power spectrum  and
bispectrum, are
\beq
\delta(\vx) = \int\dd\vk \, e^{\ii\vk\cdot\vx} \, \tdelta(\vk) ,
\label{tdelta}
\eeq
\beq
\lag \tdelta(\vk_1) \tdelta(\vk_2) \rag = \delta_D(\vk_1+\vk_2) P(k_1) ,
\label{Pkdef}
\eeq
and
\beq
\lag \tdelta(\vk_1) \tdelta(\vk_2) \tdelta(\vk_3) \rag =
\delta_D(\vk_1+\vk_2+\vk_3) \, B(k_1,k_2,k_3) .
\label{Bkdef}
\eeq
In this section we recall the model that we use to describe the 3D matter distribution,
which is presented in greater detail in \cite{Valageas2011d,Valageas2011e}.
It combines systematic perturbation theory, which governs large scales, with a  
phenomenological halo model, which governs small scales. This is achieved by
writing the 3D matter power spectrum as
\beq
P_{\rm 2H+1H}(k) = P_{\rm 2H}(k) + P_{\rm 1H}(k) ,
\label{Pk-halo}
\eeq
as in the usual halo model \citep{Cooray2002}. The main improvements with respect
to previous works are that the ``2-halo'' term is obtained from a perturbative 
resummation scheme, which is complete up to one-loop order (and includes partial
resummations of diagrams at all higher orders), while the ``1-halo'' term includes a 
``counterterm'' associated with mass conservation that ensures its
well-behaved asymptote at low $k$.

More precisely, the 2-halo contribution is written as
\beq
P_{\rm 2H}(k) = F_{\rm 2H}(2\pi/k) \, P_{\rm pert}(k) ,
\label{Pk-2H}
\eeq
where $F_{\rm 2H}(q)$ is the fraction of pairs, with initial (i.e. Lagrangian) separation
$q$, that belong to two distinct halos, and $P_{\rm pert}(k)$ is the matter power
spectrum obtained by perturbation theory. As discussed in \cite{Valageas2011d},
it is not possible to use standard perturbation theory (beyond linear order)
for $P_{\rm pert}(k)$, unless one adds a nonperturbative ad-hoc cutoff, because
this would yield a term that keeps growing at high $k$ and prevents
a good agreement with numerical simulations. A natural cure to this problem is to
use a resummation scheme that remains well-behaved at high $k$.
As in \cite{Valageas2011d}, we use the ``direct steepest-descent'' resummation
developed in \cite{Valageas2007,Valageas2008}, going to ``one-loop'' order.
This provides a perturbative term $P_{\rm pert}(k)$ that is consistent with
standard perturbation theory up to one-loop order (i.e. up to order $P_L^2$)
while keeping close to $P_L$ at high $k$ (thanks to the partial resummation of higher
orders).
The prefactor $F_{\rm 2H}(2\pi/k)$ is equal to unity at all orders of perturbation
theory, since it also writes as
\beq
F_{\rm 2H}(q) = 1-F_{\rm 1H}(q) ,
\label{F2H}
\eeq
where $F_{\rm 1H}$ is the fraction of pairs of initial separation $q$ that belong to
a single halo, and $F_{\rm 1H}=0$ at all orders of perturbation theory
(because this is a nonperturbative quantity that behaves as
$e^{-\delta_L^2/\sigma(q)^2}$ for Gaussian initial conditions).

Next, the 1-halo contribution is written as
\beq
P_{\rm 1H}(k) = \int_0^{\infty} \frac{\dd\nu}{\nu} f(\nu) 
\frac{M}{\rhob (2\pi)^3} \left(  \tu_M(k)^2 - \tW(k q_M)^2 \right) ,
\label{Pk-1H}
\eeq
where $f(\nu)$ defines the halo mass function through \citep{Press1974}
\beq
n(M) \dd M = \frac{\rhob}{M} f(\nu) \frac{\dd\nu}{\nu} , \;\; \mbox{with} \;\;
\nu = \frac{\deltaLc}{\sigma(q_M)} .
\label{nM-def}
\eeq
Here, $\sigma(q_M)$ is the rms linear density contrast at mass $M$, or
Lagrangian radius $q_M$, with
\beq
M= \rhob \frac{4\pi}{3} q_M^3 ,
\label{Mq}
\eeq
and
\beq
\sigma^2(q) = 4\pi \int_0^{\infty} \dd k \, k^2 P_L(k) \tW(kq)^2 ,
\label{sigma2-def}
\eeq
where $\tW(kq)$ is the Fourier transform of the top-hat of radius $q$, defined as
\beq
\tW(kq) = \int_V\frac{\dd \vq}{V} \, e^{\ii \vk\cdot\vq}
= 3 \, \frac{\sin(kq)-kq\cos(kq)}{(kq)^3} .
\label{tWdef}
\eeq
We use the scaling function $f(\nu)$ from \cite{Valageas2009} with the threshold
$\deltaLc$ associated with halos defined by a nonlinear density contrast of $200$.
In Eq.(\ref{Pk-1H}) $\tu_M(k)$ is the Fourier transform of the density profile
$\rho_M(x)$ of halos of mass $M$,
\beq
\tu_M(k) = \frac{\int\dd\vx \, e^{-\ii\vk\cdot\vx} \rho_M(x)}{\int\dd\vx 
\, \rho_M(x)} = \frac{1}{M} \int\dd\vx \, e^{-\ii\vk\cdot\vx} \rho_M(x) ,
\label{uM-k-def}
\eeq
whereas $\tW(k q_M)$ was defined in Eq.(\ref{tWdef}).
We use the usual ``NFW'' halo profile \citep{Navarro1997} with the mass-concentration
relation from \cite{Valageas2011d}.
The counterterm $\tW^2$ in Eq.(\ref{Pk-1H}) ensures that the 1-halo contribution
decays as $P_{\rm 1H}(k) \propto k^2$ at low $k$, so that the total power
(\ref{Pk-halo}) goes to the linear power for CDM cosmologies.
This follows from the conservation of matter and the fact that halo formation 
corresponds to small-scale redistribution of matter \citep{Peebles1974}.
In fact, taking also momentum conservation into account would yield an even
steeper $k^4$ tail \citep{Peebles1974} but the form (\ref{Pk-1H}) is sufficient
for practical purposes.

Thus, this framework ensures that the power spectrum (\ref{Pk-halo}) is consistent
with perturbation theory (up to the order of truncation of the resummation scheme)
while going smoothly to the 1-halo power at high $k$.
A more detailed derivation of Eqs.(\ref{Pk-2H}) and (\ref{Pk-1H}) is given in 
\cite{Valageas2011d}, through a Lagrangian point of view.

As described in \cite{Valageas2011e}, the 3D matter bispectrum is obtained in
a similar fashion as the sum of 3-halo, 2-halo and 1-halo terms,
\beqa
B(k_1,k_2,k_3) & = & B_{\rm 3H}(k_1,k_2,k_3) + B_{\rm 2H}(k_1,k_2,k_3) \nonumber \\
&& + B_{\rm 1H}(k_1,k_2,k_3) ,
\label{Bk-halo}
\eeqa
with
\beq
B_{3\rm H}(k_1,k_2,k_3) = B_{\rm pert}(k_1,k_2,k_3) ,
\label{B-3H}
\eeq
\beqa
B_{2\rm H}(k_1,k_2,k_3) & = &  P_L(k_1) \int \frac{\dd\nu}{\nu}
\frac{M}{\rhob(2\pi)^3} \, f(\nu)  \nonumber \\
&& \hspace{-1.cm} \times 
\prod_{j=2}^3 \left( \tu_M(k_j)  - \tW(k_jq_M) \right) + 2 \, {\rm cyc.} ,
\label{B-2H}
\eeqa
where ``2 cyc.'' stands for two terms obtained by circular permutations over
$\{k_1,k_2,k_3\}$, and
\beqa
B_{1\rm H}(k_1,k_2,k_3) & = & \int \frac{\dd\nu}{\nu} \, f(\nu) \,
\left( \frac{M}{\rhob(2\pi)^3} \right)^{\!2} \nonumber \\
&& \times \prod_{j=1}^3 \left( \tu_M(k_j)  - \tW(k_jq_M) \right) .
\label{B-1H}
\eeqa
Again, the counterterms $\tW$ in Eqs.(\ref{B-2H}) and (\ref{B-1H}) ensure that
the 2-halo and 1-halo contributions decay on large scales so that the bispectrum
goes to the perturbative prediction $B_{\rm pert}$.
However, as found in \cite{Valageas2011e} and contrary to the situation encountered
for the power spectrum, the standard one-loop perturbation theory prediction for
$B_{\rm pert}$ is well-behaved at high $k$ (i.e. it is significantly smaller than the
1-halo contribution) and it is more accurate than two alternative resummations that
have been investigated.
Therefore, we simply use the standard one-loop perturbation theory prediction for
$B_{\rm pert}$.

While Eq.(\ref{Bk-halo}) yields a
reasonably good match to numerical simulations ($\sim 10\%$) over all scales
for the bispectrum, 
Eq.(\ref{Pk-halo}) significantly underestimates the power spectrum
on the transition scales (by $\sim 20\%$) even though it gives a good accuracy
on larger scales ($\sim 1\%$) and smaller scales ($\sim 10\%$).
In \cite{Valageas2011e} we devised a simple geometrical modification to
Eq.(\ref{Pk-halo}), $P_{\rm tang}(k)$, such that 
\beq
P_{\rm tang}(k) = P_{\rm 2H+1H}(k) \;\; \mbox{for} \;\; k \leq k_- \;\; \mbox{or} \;\; k \geq k_+' ,
\label{P-tang-1}
\eeq
and
\beq
\log\left[ \frac{P_{\rm tang}(k)}{P_{\rm tang}(k_-)}\right] = 
\frac{\log(k/k_-)}{\log(k_+'/k_-)} \log\left[\frac{P_{\rm tang}(k_+')}{P_{\rm tang}(k_-)}\right]
\label{P-tang-2}
\eeq
for $k_-<k<k_+'$.
In other words, $P_{\rm tang}$ is identical to $P_{\rm 2H+1H}$ of Eq.(\ref{Pk-2H})
outside of the transition range $[k_-,k_+']$, and we interpolate by a straight line
in the plane $\{\log k, \log P\}$ over the interval $[k_-,k_+']$. 
This partly cures the underestimate of the power spectrum in this range, while keeping
the perturbative and 1-halo behaviors on large and small scales. 
As explained in \cite{Valageas2011e}, the wavenumbers $k_-$ and $k_+'$ are determined
by the constraint that the reduced equilateral bispectrum, $Q_{\rm eq}(k)=B(k,k,k)/[3P(k)^2]$,
is monotonically increasing. Indeed, it happens that Eq.(\ref{Pk-halo}) underestimates the
actual power spectrum on transition scales, which leads to a spurious peak for $Q_{\rm eq}(k)$.
Then, from the shape of this naive prediction for $Q_{\rm eq}(k)$ we automatically detect the
range $[k_-,k_+']$ where the model is not sufficienly accurate (because higher-order
perturbative contributions come into play or the decomposition over 2-halo and 1-halo terms
is too simple to accurately describe regions such as outer shells and filamentary
branches of collapsing halos that have not yet relaxed).
This typically gives $k_-=0.4 h$Mpc$^{-1}$ and $k_+'=3 h$Mpc$^{-1}$ at $z=1$.
The main point here is that the range $[k_-,k_+']$ is automatically determined from the shape
of $P_{\rm 2H+1H}(k)$ and $B(k,k,k)$, and it evolves with redshift and with the initial
power. 
We refer the reader to \cite{Valageas2011e} for further details.

Thus, Eqs.(\ref{Bk-halo})-(\ref{B-1H}) and (\ref{P-tang-1})-(\ref{P-tang-2}) define
the analytic model that we investigate in this paper for the 3D matter bispectrum
and power spectrum.

\subsection{2D weak-lensing power spectrum and bispectrum}
\label{lensing-power}

Using Born's approximation, the weak-lensing convergence $\kappa(\vtheta)$ can be
written as the integral of the density contrast along the line of sight
\citep{Bartelmann2001,Munshi2008},
\beq
\kappa(\vtheta) = \int_0^{\chi_s} \dd\chi \, w(\chi,\chi_s) \, \delta(\chi,\cD\vtheta) ,
\label{kappa-def}
\eeq
where $\chi$ and $\cD$ are the radial and angular comoving distances,
\beq
w(\chi,\chi_s) = \frac{3\Omega_{\rm m} H_0^2 \cD(\chi) \cD(\chi_s-\chi)}{2c^2\cD(\chi_s)} (1+z) ,
\label{w-def}
\eeq
and $z_s$ is the redshift of the source (in this article we only consider the case where
all sources are located at a single redshift, to simplify the comparisons with numerical
simulations and the dependence on the source redshift).
Then, using a flat-sky approximation, which is valid for small angles below a few
degrees \citep{Valageas2011c}, we define its 2D Fourier transform through
\beq
\kappa(\vtheta) = \int \dd\vell \, e^{\ii\vell\cdot\vtheta} \, \tkappa(\vell) .
\label{kappa-l-def}
\eeq
As in 3D, we define the 2D convergence power spectrum and bispectrum as
\beq
\lag \tkappa(\vell_1) \tkappa(\vell_2) \rag = \delta_D(\vell_1+\vell_2) \Pkappa(\ell_1) ,
\label{Pkappa-def}
\eeq
and
\beq
\lag \tkappa(\vell_1) \tkappa(\vell_2) \tkappa(\vell_3) \rag =
\delta_D(\vell_1+\vell_2+\vell_3) \, \Bkappa(\ell_1,\ell_2,\ell_3) .
\label{Bkappa-def}
\eeq
From Eq.(\ref{kappa-def}) one obtains at once, using Limber's approximation
\citep{Limber1953,Kaiser1992,Bartelmann2001,Munshi2008},
\beq
\Pkappa(\ell)= 2\pi \int_0^{\chi_s} \! \dd\chi \, \frac{w^2}{\cD^2} \, P(\ell/\cD;z) ,
\label{Pkappa-P}
\eeq
\beq
\Bkappa(\ell_1,\ell_2,\ell_3)= (2\pi)^2 \int_0^{\chi_s} \!\! \dd\chi \,
\frac{w^3}{\cD^4} \, B(\ell_1/\cD,\ell_2/\cD,\ell_3/\cD;z) .
\label{Bkappa-B}
\eeq
This provides the weak-lensing convergence power spectrum and bispectrum
from our model described in Sect.~\ref{3D-power} for the 3D matter density
power spectrum and bispectrum, through a simple integration over the radial 
coordinate up to the source plane.

\section{Numerical simulations}
\label{Numerical}

To obtain accurate predictions of the convergence power
spectrum and bispectrum we must perform the ray-tracing simulations
through large-volume, high-resolution $N$-body simulations of structure
formation~\citep{Jain2000,Hamana2001a,Hilbert2009,Sato2009}.
To perform the $N$-body simulations, we use a modified version of the {\tt
Gadget-2} code~\citep{Springel2005}.
The ray-tracing simulations are constructed from $2\times 200$
realizations of $N$-body simulations with cubic 240 and 480$h^{-1}$Mpc
on a side, respectively.
We employ 256$^3$ particles for each simulation.
For our fiducial cosmology, we adopt the standard $\Lambda$CDM cosmology
with matter fraction $\Omega_{\rm m}=0.238$, baryon fraction
$\Omega_{\rm b}=0.0416$, dark energy fraction $\Omega_{\rm de}=0.762$ with
the equation of state parameters $w_0=-1$ and $w_a=0$, spectral index
$n_s=0.958$, normalization $A_{\rm s}=2.35\times 10^{-9}$, and Hubble
parameter $h=0.732$, which are consistent with the WMAP 3-year results~\citep{Spergel2007}. 
This fiducial cosmology gives for the variance
of the density fluctuations in a sphere of radius 8$h^{-1}$Mpc the
normalization $\sigma_8=0.759$.
We consider source redshifts at either $z_s=0.6$, 1.0, or 1.5.
Using ray-tracing simulations we generate 1000 realizations of
5$^{\circ}\times 5^{\circ}$ convergence maps for each source redshift.

In addition to the fiducial cosmology case, we performed
ray-tracing simulations for several slightly different cosmologies.
We vary each of the following cosmological parameters:
$A_s$, $n_s$, the cold dark matter density $\Omega_{\rm c}h^2$,
$\Omega_{\rm de}$, and $w_0$ by $\pm 10\%$, respectively.
In varying the parameters, we keep the flatness of the universe and the physical baryon
density $\Omega_{\rm b}h^2$ to be unchanged.
Therefore the three parameters, $h$, $\Omega_{\rm m}$, and $\Omega_{\rm b}$, 
are varied simultaneously. 
For each of  these 10 different
cosmologies, we obtain 40 realizations of convergence fields for each
of the three source redshifts.
Details of the methods used for the ray-tracing simulations can be found in
\citet{Sato2009}.
All convergence maps used in this paper are the same as those used in
\citet{Seo2011}, but for varied cosmologies we use three orthogonal
projection axes instead of using only one projection axis to increase
the independence.

In Sects.~\ref{Convergence-power-spectrum}-\ref{contributions},
we use the maps for the fiducial cosmology while in
Sect.~\ref{Cosmology} we use those for the varied cosmologies to
investigate the robustness of our model.
In Sect.~\ref{Cosmology} we show the results for six cases,
varying $n_s$, $\Omega_{\rm c}h^2$, and $w_0$ by $\pm 10\%$, which we
denote as $n_s\pm$, $\Omega_{\rm c}h^2\pm$, and $w_0\pm$.
The exact values for these cosmological parameters are listed in
Table~\ref{Table_cosmo} in App.~\ref{alternatives}.

The binned convergence power spectrum and bispectrum are measured as follows.
We first apply Fast Fourier Transformation to each of the convergence fields to obtain $\tilde{\kappa}(\vell)$.
We then bin the data into logarithmically-equal bins in $\ell$, whose widths are set as $\Delta \ln \ell=0.3$ for
the power spectrum, and $\Delta \ln \ell = \ln 2 /2 \approx 0.35$ for the bispectrum, respectively.
The power spectrum and the bispectrum are obtained by simply averaging the products of modes:
\begin{eqnarray}
\hat{\Pkappa}(\ell)&=&\frac{1}{N_{\ell}}\sum_{|\vell|\in\ell}|\tilde{\kappa}(\vell)|^2,
\label{Pkappa-Est}\\
\hat{\Bkappa}(\ell_1,\ell_2,\ell_3)&=&\frac{1}{N_{\ell_1,\ell_2,\ell_3}}\sum_{|\vell_i|\in\ell_i}{\rm Re}\left[\tilde{\kappa}(\vell_1)\tilde{\kappa}(\vell_2)\tilde{\kappa}(\vell_3)\right],
\end{eqnarray}
where Re[...] denotes the real part of a complex number, and the summation runs over modes $\vell$ ($\vell_i$, $i=1,2,3$) 
which fall into bin $\ell$ ($\ell_i$) for the power (bi-) spectrum. 
In the above, $N_{\ell}$ is the number of modes taken
for the summation and is given by $N_{\ell}\approx A_{\rm shell}\Omega_{\rm s}/(2\pi)^2$ where $A_{\rm shell}$ is the area 
of two-dimensional shell around the bin $\ell$ and can be given as $A_{\rm shell}\approx
2\pi\ell\Delta\ell+\pi(\Delta\ell)^2$, and $\Omega_{\rm s}$ is the survey area.
Similarly, the factor $N_{\ell_1,\ell_2,\ell_3}$, which appears in the estimator of the bispectrum, denotes the number of triangles in $\ell$ space.
We then average the measured spectra over $1000$ random realizations to obtain our final estimates of $\Pkappa$ and $\Bkappa$. 
The statistical uncertainty of our estimates are also estimated from the variance of $\hat\Pkappa$ and $\hat\Bkappa$ over 1000 realizations.
We plot the 3-$\sigma$ uncertainty as error bars in what follows.

\section{Convergence power spectrum}
\label{Convergence-power-spectrum}

\begin{figure*}
\begin{center}
\epsfxsize=6.1 cm \epsfysize=5.4 cm {\epsfbox{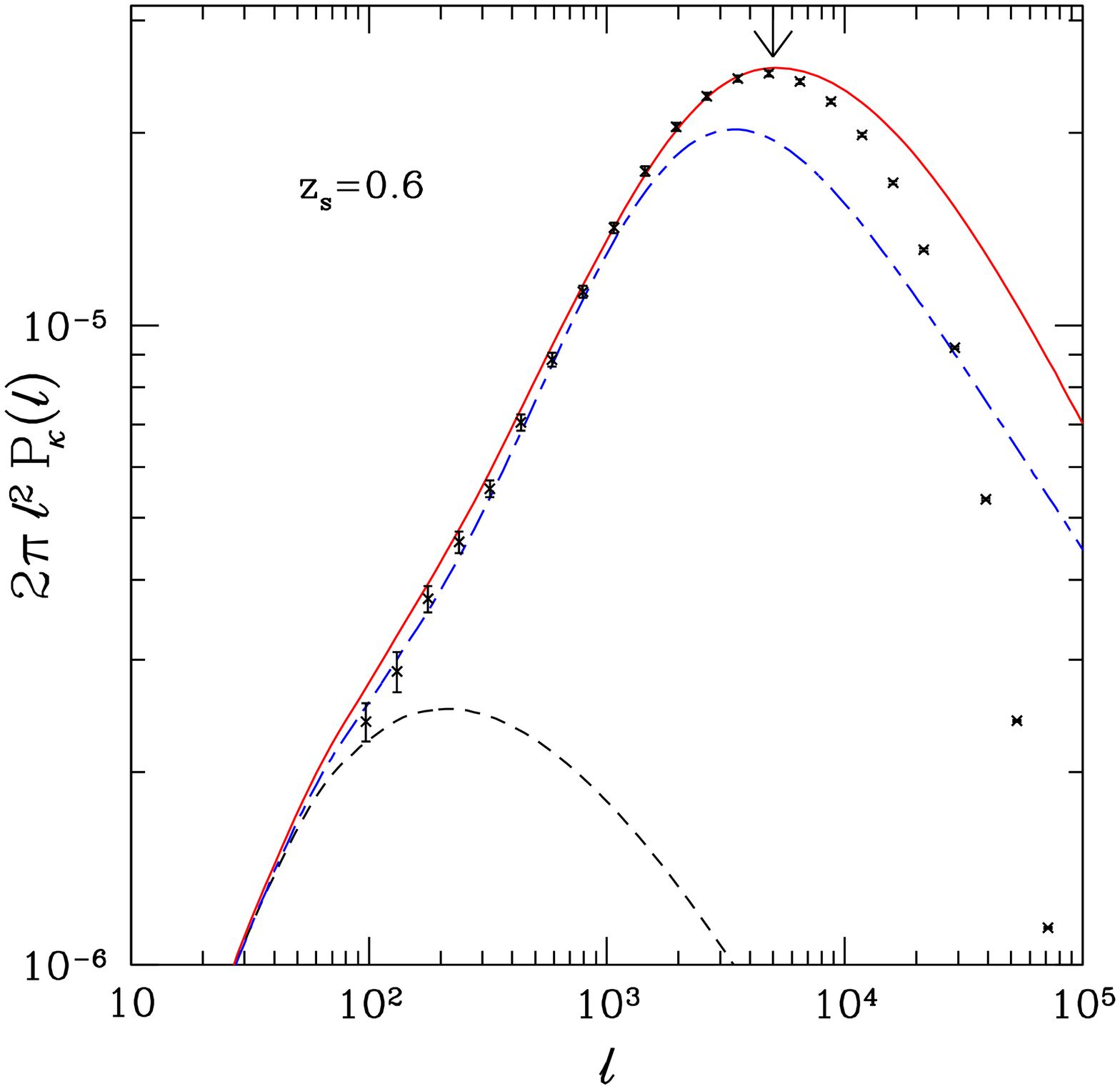}}
\epsfxsize=6.05 cm \epsfysize=5.4 cm {\epsfbox{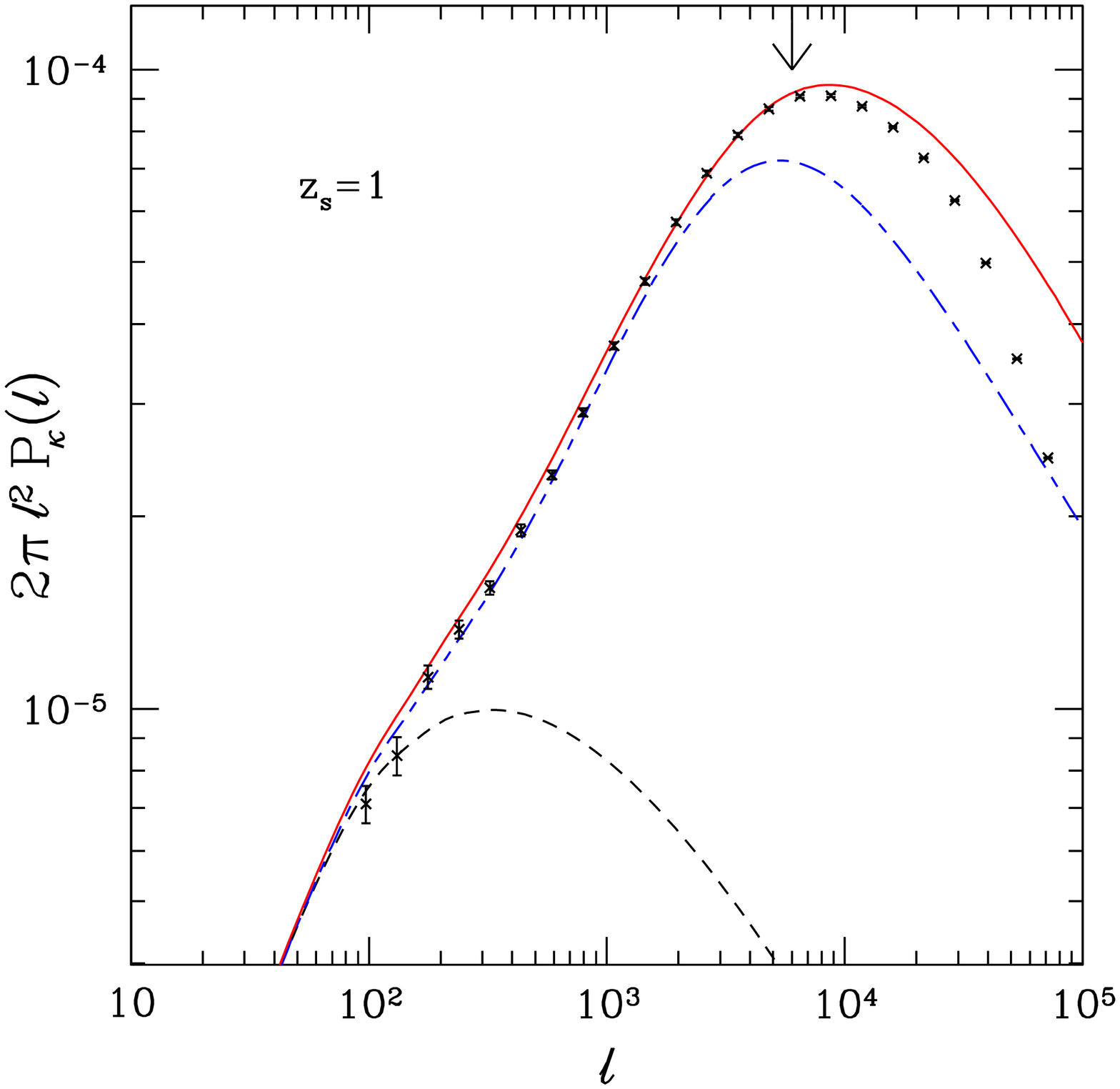}}
\epsfxsize=6.05 cm \epsfysize=5.4 cm {\epsfbox{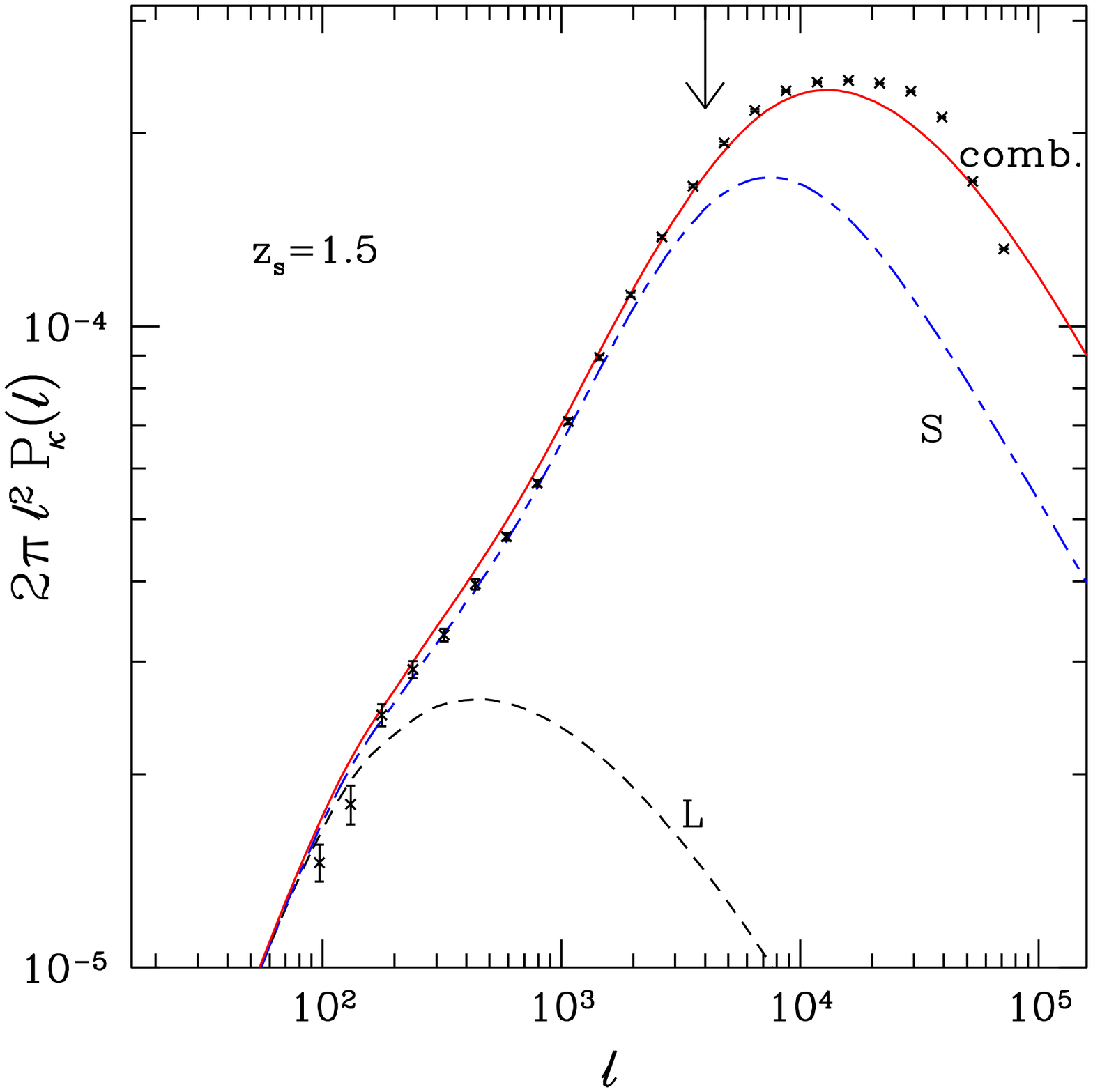}}
\end{center}
\caption{Convergence power spectrum for sources at redshifts $z_s=0.6, 1$, and
$1.5$. The points are the results from numerical simulations with $3-\sigma$
error bars.
The low black dashed line ``L'' is the linear power, the middle blue dot-dashed line
``S'' is the result from the ``halo-fit'' of \cite{Smith2003}, and the upper red solid
line ``comb.'' is the result from our model, which combines 1-loop perturbation 
theory with a halo model. The vertical arrow close to the peak shows the
scale up to which the simulation result is valid within $5\%$.}
\label{fig_Pl}
\end{figure*}

We now compare our results for the weak-lensing power spectrum with numerical
simulations. In addition, we also consider the predictions obtained from the
popular ``halo-fit'' fitting function for the 3D density power spectrum given
in \cite{Smith2003}, to estimate the gains that can be reached by using a somewhat
more systematic approach.

We show our results for the convergence power spectrum, $\Pkappa(\ell)$, in
Fig.~\ref{fig_Pl}.
For reference we also plot the linear power and we can check that both simulations and 
analytical models recover the linear regime on large scales. 
The numerical error bars increase on large scales because of the finite size of the
simulation box. On small scales the numerical error is dominated by systematic
effects, due to the finite resolution, that lead to an underestimate of the small-scale
power (as clearly shown by the sharp decline at very high $\ell$).
For each source redshift we estimated the scale $\ell$ up to which the simulations
have an accuracy of better than $5\%$ by comparing with higher-resolution simulations
(with $512^3$ particles instead of $256^3$). This scale is shown by the vertical
arrow in Fig.~\ref{fig_Pl} and we can check that our model indeed agrees
with the numerical simulations up to this multipole.

We recover the well-known fact that on scales that are of interest for cosmological
analysis of weak gravitational lensing, $10^2 < \ell < 10^4$, nonlinear terms
significantly contribute to or dominate the power spectrum
\citep{Bartelmann2001,Munshi2008}.
The overall shapes of the power spectra obtained from our model, described in
Sect.~\ref{Analytic}, and from the fitting formula from \cite{Smith2003}, are very
similar. This is expected since the latter ``halo-fit'' also goes to the linear power on
large scales and it is based on numerical simulations for its small-scale behavior,
while the halo model that underlies our approach on small scales also agrees
with similar simulations (in particular it involves the ``NFW'' profile and a
mass-concentration relation that are derived from simulations).
These similar shapes also confirm that the sharp falloff of the power at high $\ell$
found in the ray-tracing simulations is not physical but due to the finite resolution.

The improvements that can be expected from our approach, as compared with a
simpler fitting formula to measures of the 3D power, are that i) on large scales
we are consistent with perturbation theory up to one-loop order, and ii) on small
scales we directly use a physical halo model (instead of using some formula for the
power spectrum with free exponents  that are fitted to a set of simulations).

On large scales the error bars of our ray-tracing simulations (and their lack of 
large-scale power) are too large to clearly see the benefit of the one-loop 
perturbative terms at low $z_s$. However, the higher accuracy due to these
higher-order perturbative contributions can be seen at $z_s=1$ and $z_s=1.5$,
as will be shown more clearly in Sect.~\ref{Convergence-power-spectrum-1H}
below. (The benefit of these terms has already been shown in studies of
the 3D matter density power spectrum, especially for the accurate prediction
of the baryon acoustic oscillations, e.g.
\citet{Jeong2006,Nishimichi2007b,Nishimichi2009,Crocce2008,Matsubara2008,Taruya2009,Sato2011b,Valageas2011d,Valageas2011e}).
In the weak-lensing context, this higher accuracy could also be useful for the
analysis of future observations such as the Euclid mission \citep{Refregier2010}.

On small scales, although the agreement of our model with the simulations is not 
perfect we can see a clear improvement as compared with the fitting formula
from \cite{Smith2003}.
This is not surprising since the latter was derived from a set of older numerical simulations
with somewhat different cosmological parameters than the ones we consider in this
paper. The underestimate of the power on small scales by this ``halo-fit'' formula,
by about $30\%$ around the peak for $z_s=1$, was already noticed in 
\cite{Hilbert2009}.
The comparison of our model with the 3D power spectrum in 
\cite{Valageas2011d,Valageas2011e} was also based on simulations with different 
cosmological parameters, but through the explicit expression (\ref{Pk-1H})
we automatically take into account the dependence on cosmology of the linear
growth factor $D_+(t)$ of the density fluctuations and of the linear threshold
$\deltaLc$ 
associated with the nonlinear density contrast of $200$ \citep{Valageas2009}.
We mostly neglect the dependence on cosmology of the halo profiles and of their
mass-concentration relation. However, even though virialization processes do not
reach complete relaxation, that is, a violent relaxation ``\`a la Lynden-Bell''
\citep{Lynden-Bell1967} is not complete in this cosmological context and does not  
allow the halos to reach a statistical equilibrium that is fully independent of the
initial conditions, internal halo properties are almost universal and independent of
cosmological parameters up to a good accuracy (for realistic CDM scenarios).
Similarly, deviations of the halo mass function from ``universality'' have been detected
but are rather weak \citep{Tinker2008,More2011,Bhattacharya2011}.
This is especially true for our purpose as we integrate over the full halo mass function
(which is normalized to unity) and we take into account the dependence on cosmology
of its large-mass cutoff.
This explains why our model is able to reach a good agreement with the
numerical simulations. 

On the intermediate scales, which are sensitive to the interpolation 
(\ref{P-tang-2}), we also obtain a satisfactory match to the
numerical simulations. Within this range, around $\ell \sim 300$, the ``halo-fit'' prediction
happens to provide a similar, or in a few cases slightly better, agreement with the
simulations. These transition scales, where both the one-loop
perturbative contribution and the 1-halo term are subdominant as shown in
Fig.~\ref{fig_Pl_1H} below, are governed by the interpolation (\ref{P-tang-2}).
This means that this interpolation is not fully satisfactory and that there remains room
for improvement. However, to keep our model as simple as possible and to
remain consistent with our previous 3D studies we do not investigate here
alternative prescriptions.

We shall check in Sect.~\ref{Cosmology} below that the overall agreement found
in Fig.~\ref{fig_Pl} remains valid as we change the values of the cosmological parameters.

\section{Convergence bispectrum}
\label{Convergence-bispectrum}

\begin{figure*}
\begin{center}
\epsfxsize=6.1 cm \epsfysize=5.4 cm {\epsfbox{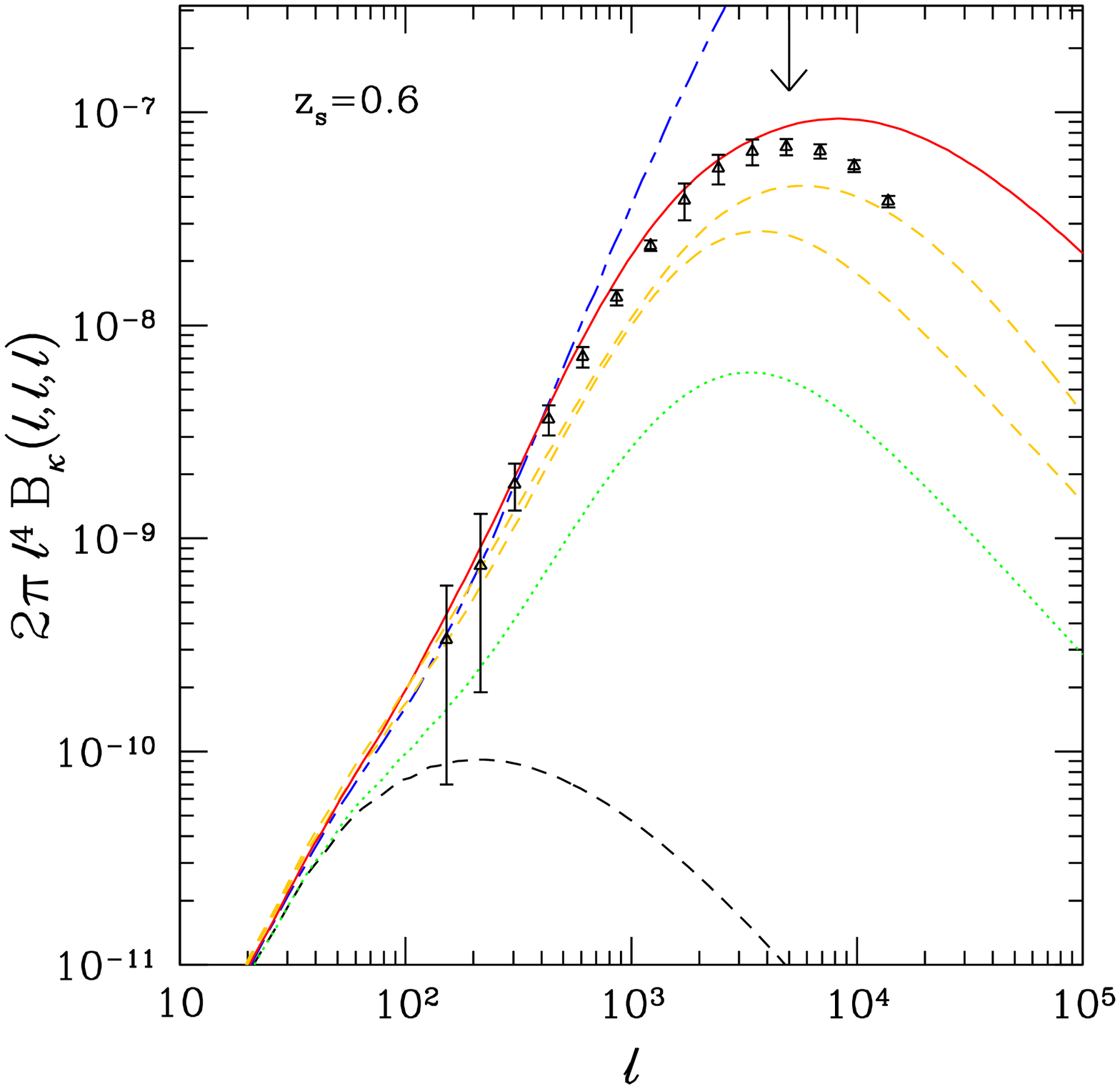}}
\epsfxsize=6.05 cm \epsfysize=5.4 cm {\epsfbox{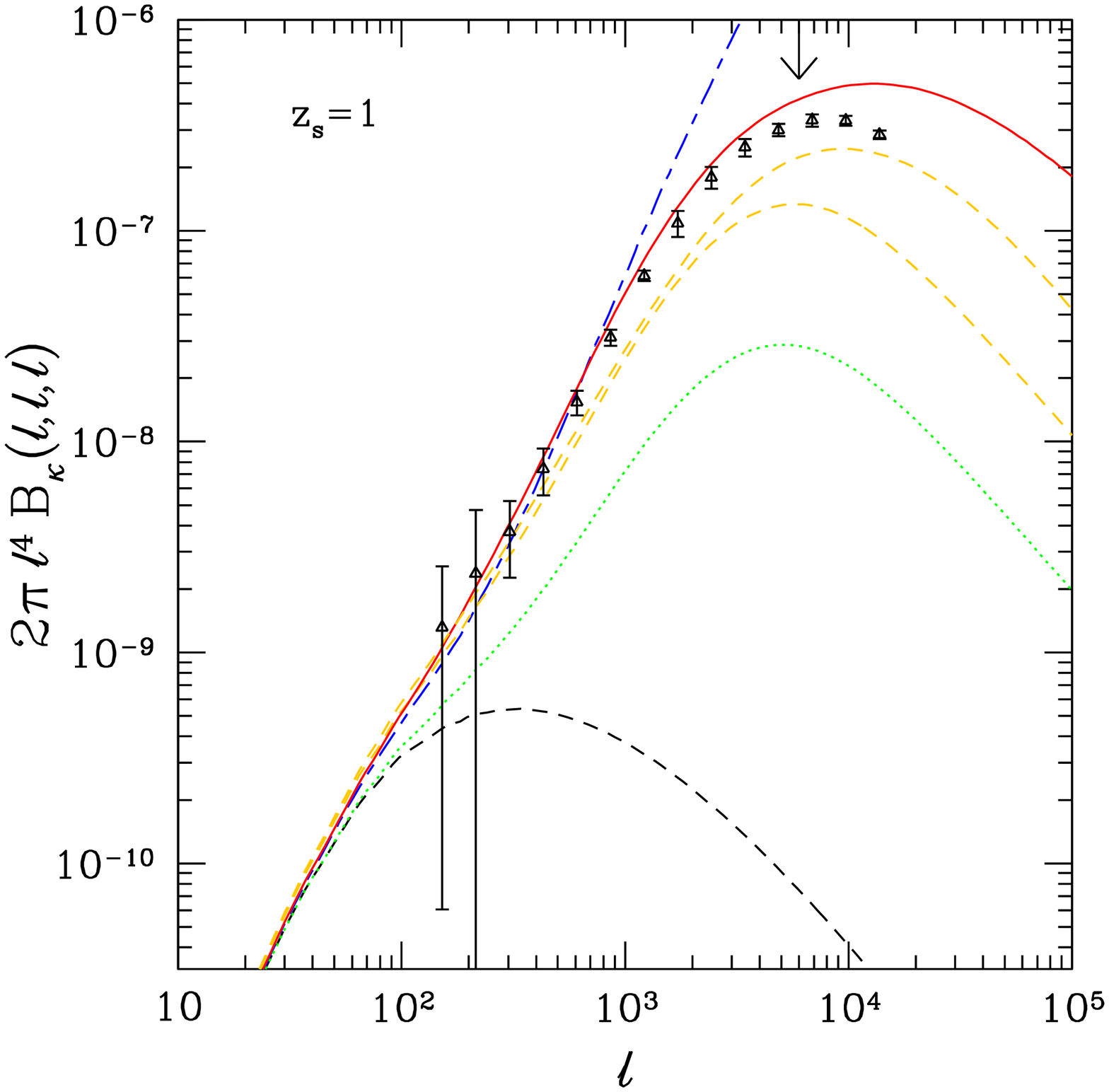}}
\epsfxsize=6.05 cm \epsfysize=5.4 cm {\epsfbox{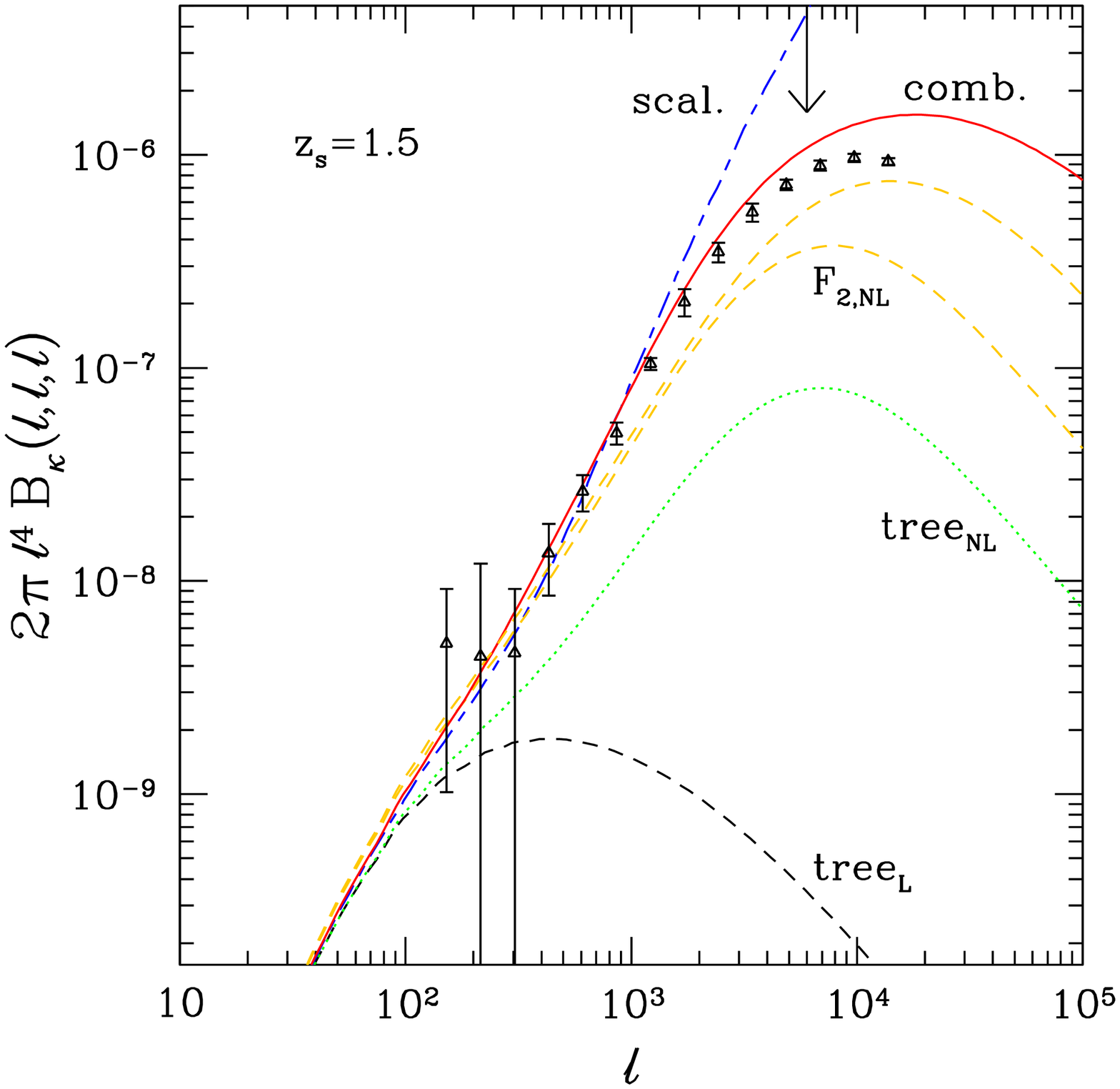}}
\end{center}
\caption{Convergence bispectrum for sources at redshifts $z_s=0.6, 1$, and
$1.5$, for equilateral triangles.
The points are the results from numerical simulations with $3-\sigma$ error bars.
The low black dashed line ``$\rm tree_L$'' is the tree-order bispectrum
(\ref{B-treeL-def}), the green dotted line ``$\rm tree_{NL}$'' is the tree-order formula
(\ref{B-treeNL-def}), where we use the nonlinear power from the ``halo-fit'' of 
\cite{Smith2003}, the two yellow dashed lines ``$F_{2,\rm NL}$'' (\ref{B-F2NL-def})
include in addition a fitting formula for the kernel $F_{2,\rm NL}$ from 
\cite{Scoccimarro2001a} and use either $P_S(k)$ (lower curve) or $P_{\rm tang}(k)$
(upper curve) for the 3D power spectrum, the upper blue dot-dashed line ``scal.'' is
the phenomenological scale transformation (\ref{B-Pan-def}) from \cite{Pan2007},
and the red solid line ``comb.'' is our combined model (\ref{Bk-halo})-(\ref{B-1H}).
The vertical arrows are at the same scale as in Fig.~\ref{fig_Pl}.}
\label{fig_Beq}
\end{figure*}

We now compare our results for the weak-lensing bispectrum with
numerical simulations. In addition, we also consider the predictions obtained from
the following four simple models, which have been used in some previous works.

We first compute the results obtained from the lowest-order (``tree-order'') prediction
from standard perturbation theory for the 3D bispectrum \citep{Bernardeau2002},
\beq
B_{\rm tree_L}(k_1,k_2,k_3) \! = 2 F_2(k_1,k_2,\mu_{12}) \, P_L(k_1) P_L(k_2)
+ 2 \, {\rm cyc.} 
\label{B-treeL-def}
\eeq
where $\mu_{12}= (\vk_1\cdot\vk_2)/(k_1 k_2)$ and
\beq
F_2(k_1,k_2,\mu_{12}) = \frac{5}{7}
+ \frac{1}{2} \left(\frac{k_1}{k_2}\!+\!\frac{k_2}{k_1}\right) \mu_{12} 
+ \frac{2}{7} \, \mu_{12}^2 .
\label{F2-def}
\eeq

Second, we consider a ``tree-nonlinear'' approximation where in Eq.(\ref{B-treeL-def})
we replace the linear 3D power $P_L(k)$ by the nonlinear power $P_S(k)$ from
\cite{Smith2003},
\beq
B_{\rm tree_{NL}}(k_1,k_2,k_3) \! = 2 F_2(k_1,k_2,\mu_{12}) \, P_S(k_1) P_S(k_2)
+ 2 \, {\rm cyc.} 
\label{B-treeNL-def}
\eeq

Third, we consider the fitting formula from \cite{Scoccimarro2001a},
\beqa
B_{F_{2,\rm NL}}(k_1,k_2,k_3) & = & 2 \, F_{2,\rm NL}(k_1,k_2,\mu_{12})  P_S(k_1) P_S(k_2) 
\nonumber \\
&& + 2 \, {\rm cyc.} 
\label{B-F2NL-def}
\eeqa
where the kernel $F_{2,\rm NL}$ is an effective kernel that interpolates from the
large-scale perturbative result (\ref{F2-def}) to a small-scale ansatz where
the angular dependence on $\mu_{12}$ vanishes. Here we shall use the
nonlinear power spectrum $P_S(k)$ from \cite{Smith2003} as well as the
nonlinear power spectrum $P_{\rm tang}(k)$ of our model,
Eqs.(\ref{P-tang-1})-(\ref{P-tang-2}).

Four, we consider the scale transformation studied in \cite{Pan2007},
following the spirit of the scale transformation introduced in \cite{Hamilton1991}
for the two-point correlation function and next in \cite{Peacock1996} for the
power spectrum,
\beqa
B_{\rm scal}(k_1,k_2,k_3) & = & w(\tk_1) w(\tk_2) w(\tk_3) 2 F_2(\tk_1,\tk_2,\tmu_{12}) 
\nonumber \\
&& \times \, P_L(\tk_1) P_L(\tk_2) + 2 \, {\rm cyc.} 
\label{B-Pan-def}
\eeqa
with for $i=1,2,3$,
\beq
\tk_i = [1+\Delta^2_S(k_i)]^{-1/3} k_i , \;\;\; 
w(\tk_i)= \sqrt{P_S(\tk_i)/P_L(\tk_i)} .
\eeq

\subsection{Equilateral triangles}
\label{Equilateral}

\begin{figure*}
\begin{center}
\epsfxsize=6.1 cm \epsfysize=5.4 cm {\epsfbox{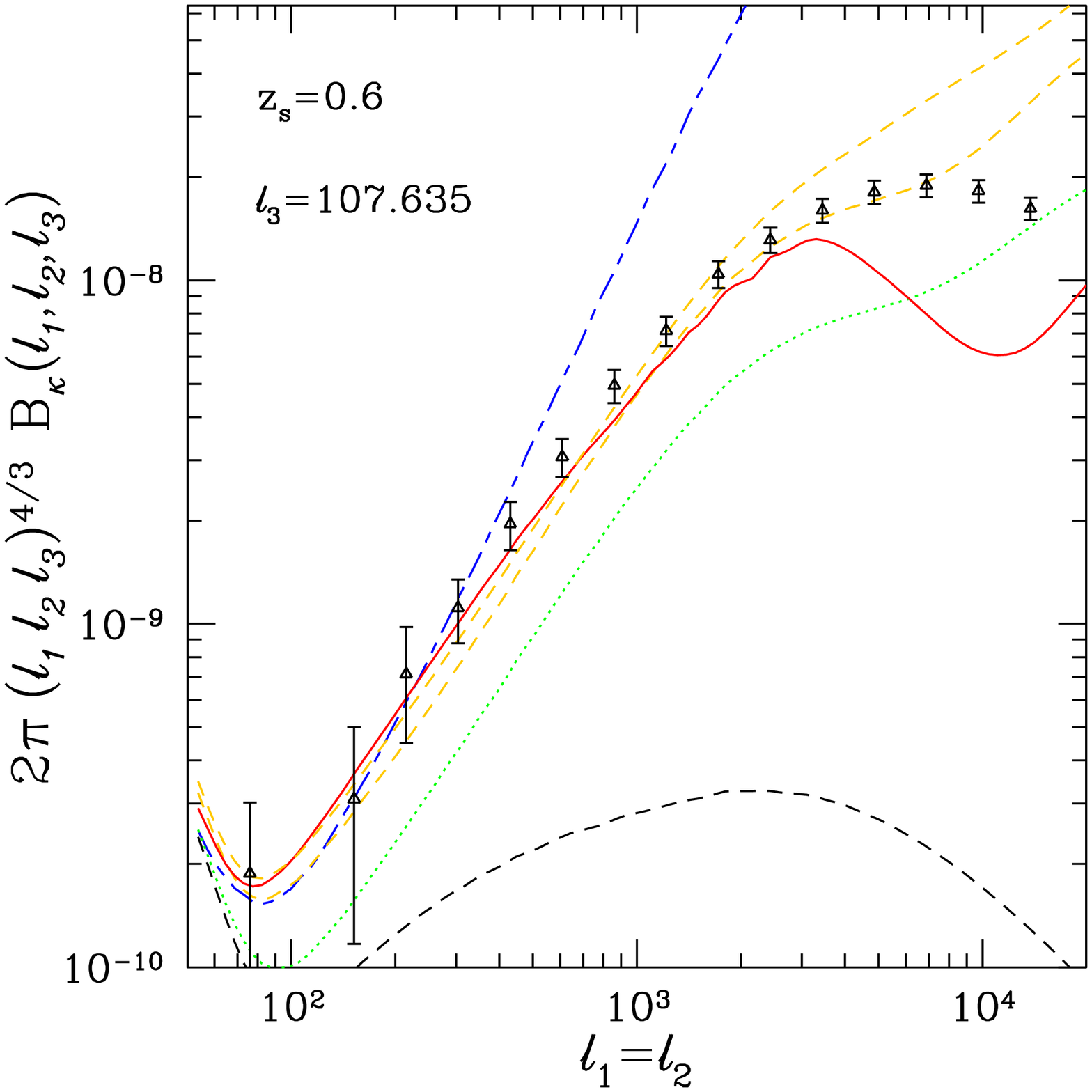}}
\epsfxsize=6.05 cm \epsfysize=5.4 cm {\epsfbox{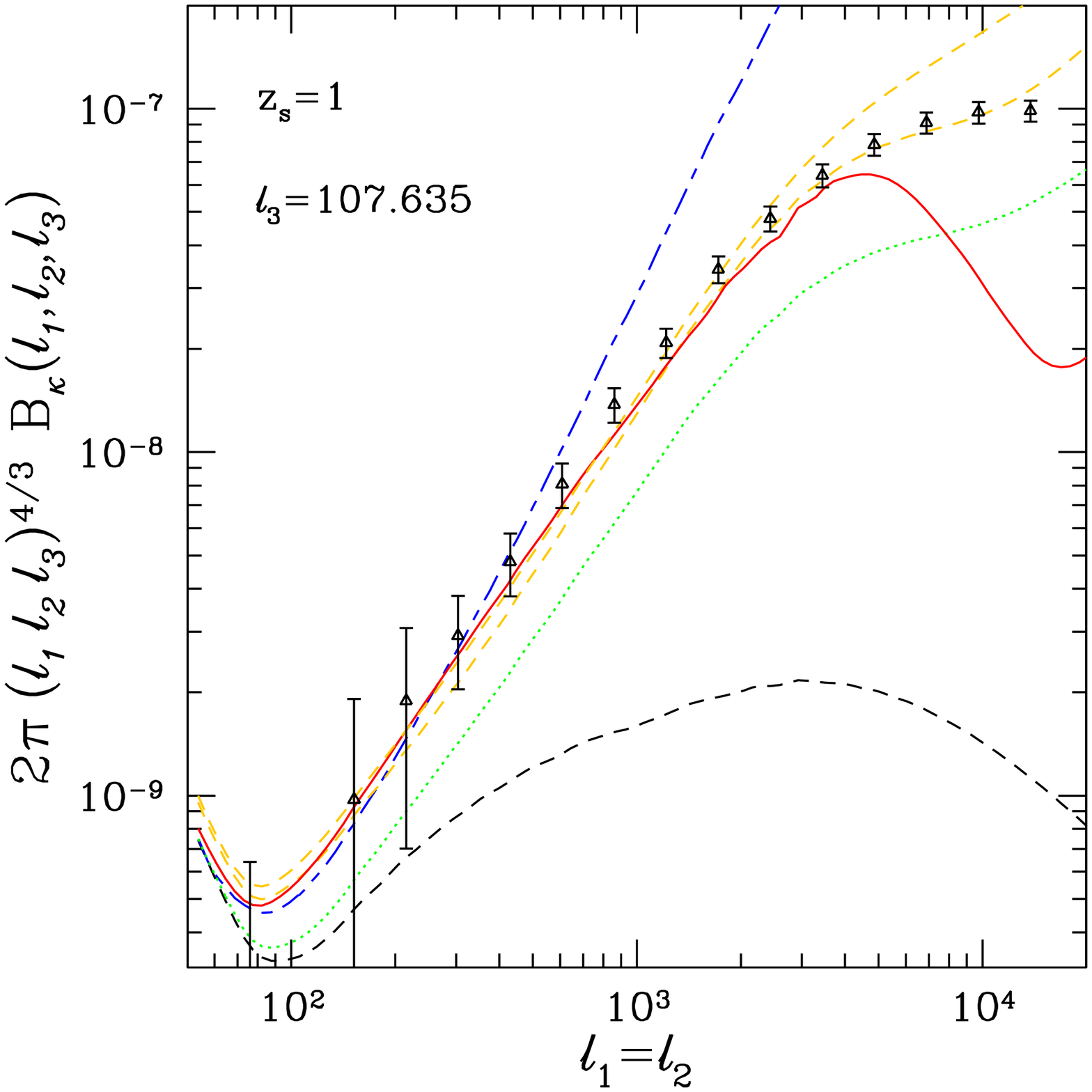}}
\epsfxsize=6.05 cm \epsfysize=5.4 cm {\epsfbox{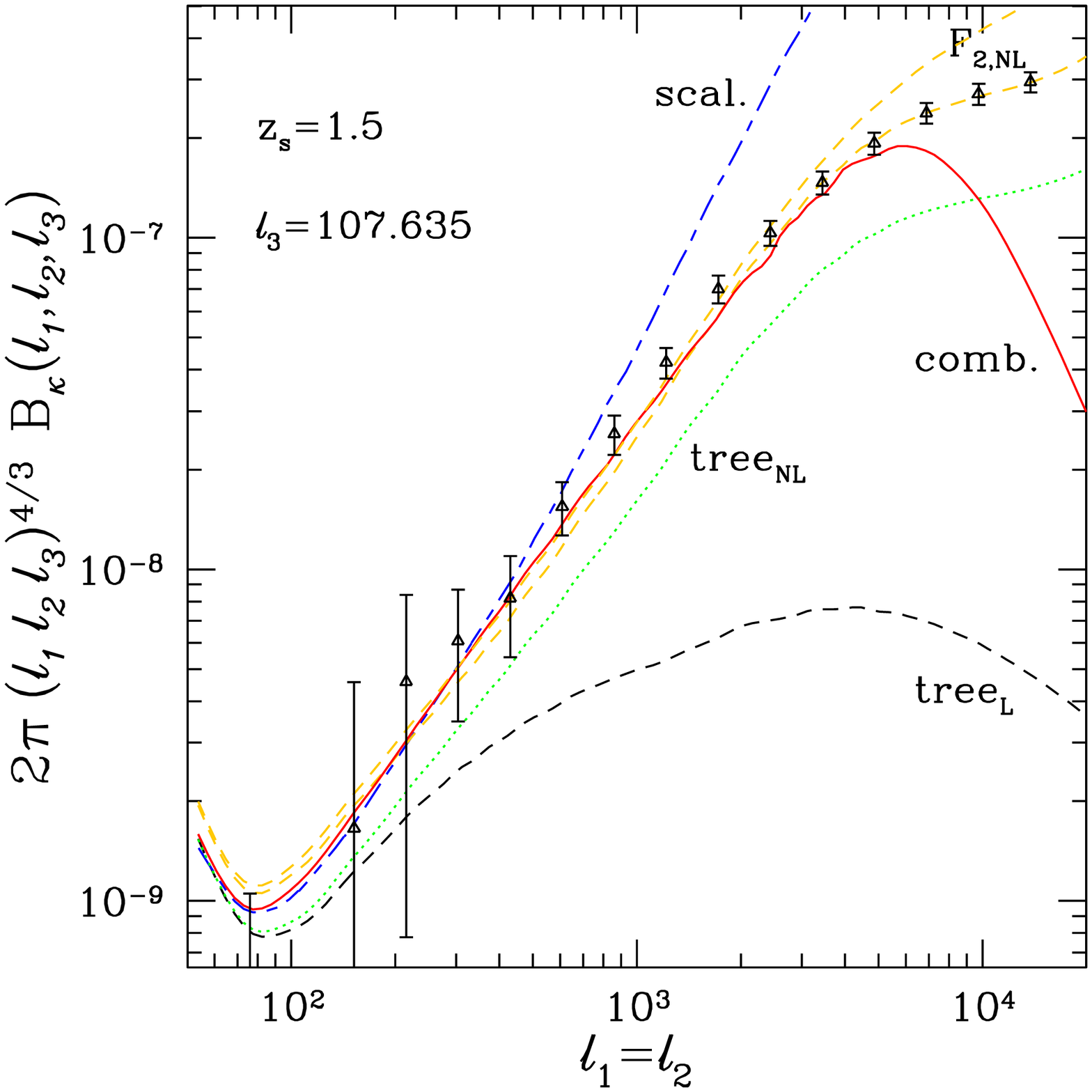}}\\
\epsfxsize=6.1 cm \epsfysize=5.4 cm {\epsfbox{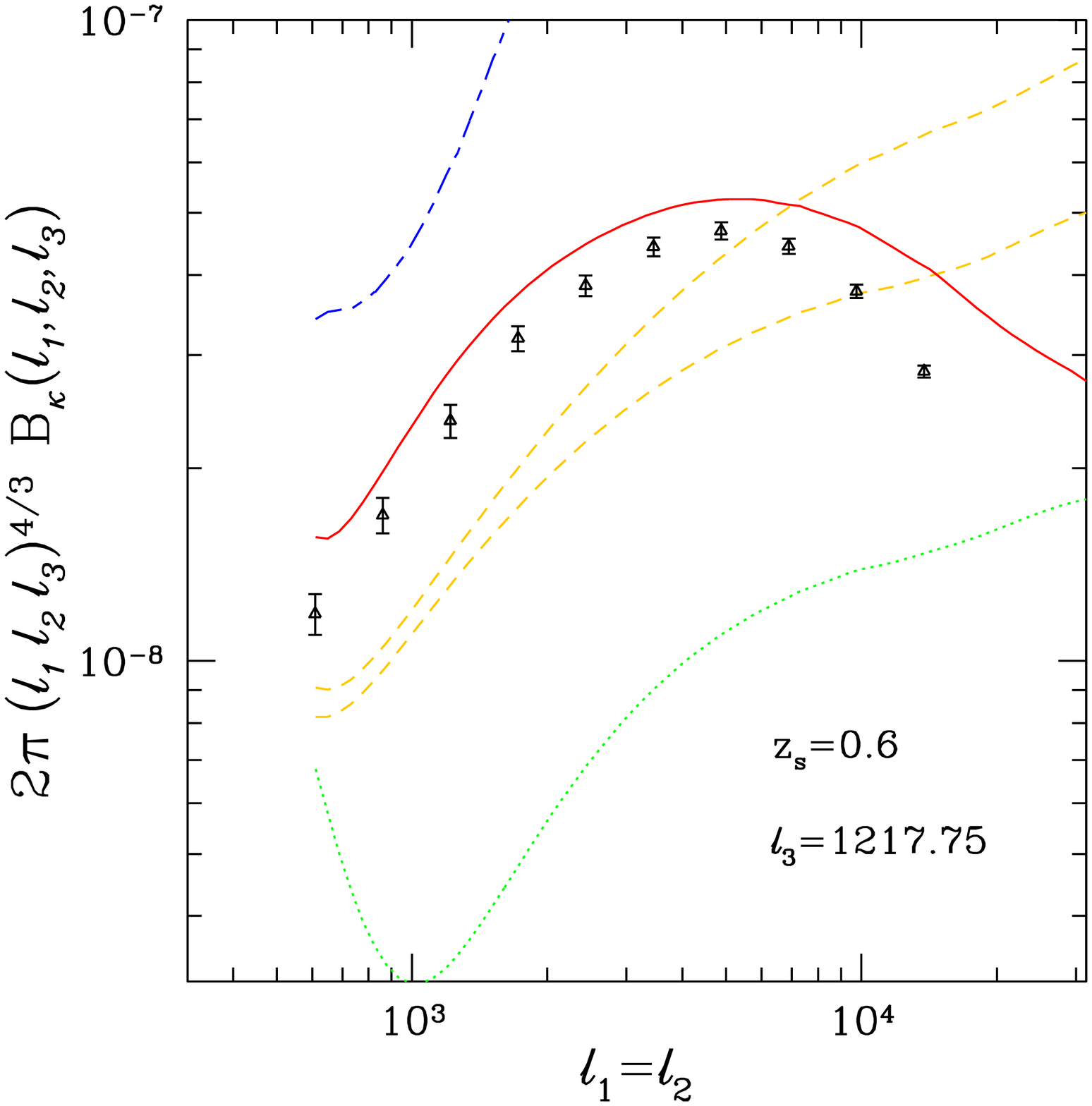}}
\epsfxsize=6.05 cm \epsfysize=5.4 cm {\epsfbox{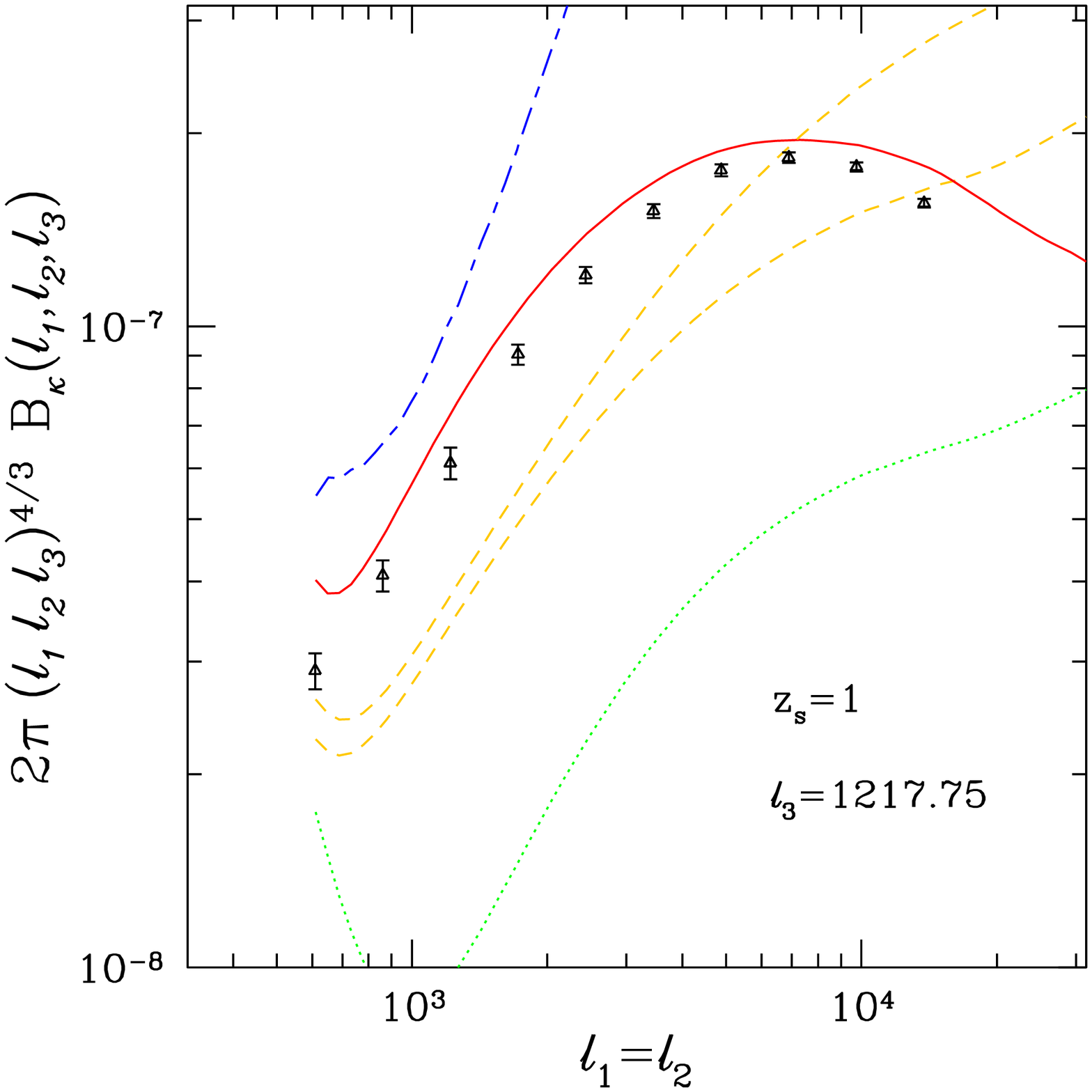}}
\epsfxsize=6.05 cm \epsfysize=5.4 cm {\epsfbox{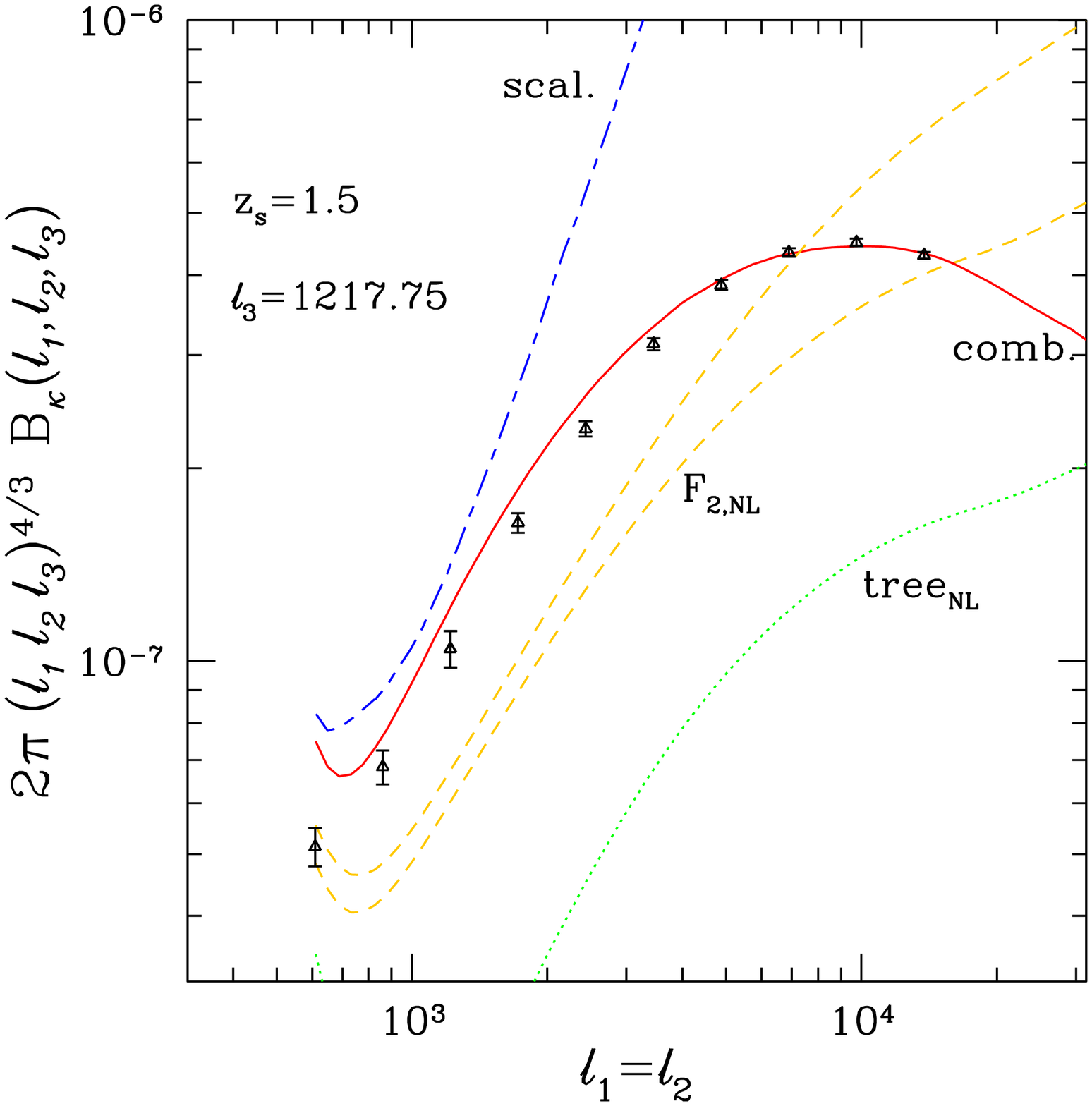}}\\
\epsfxsize=6.1 cm \epsfysize=5.4 cm {\epsfbox{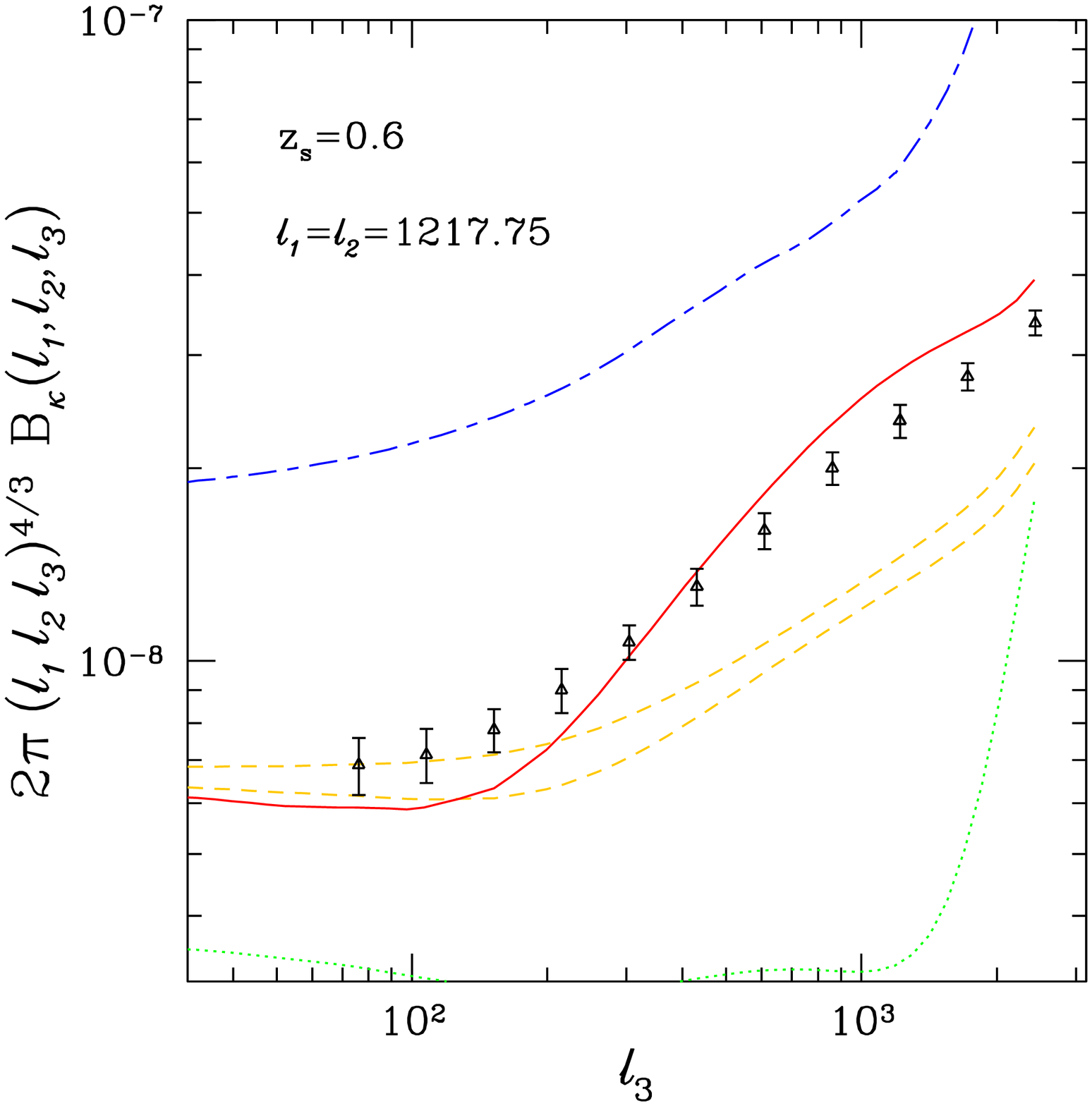}}
\epsfxsize=6.05 cm \epsfysize=5.4 cm {\epsfbox{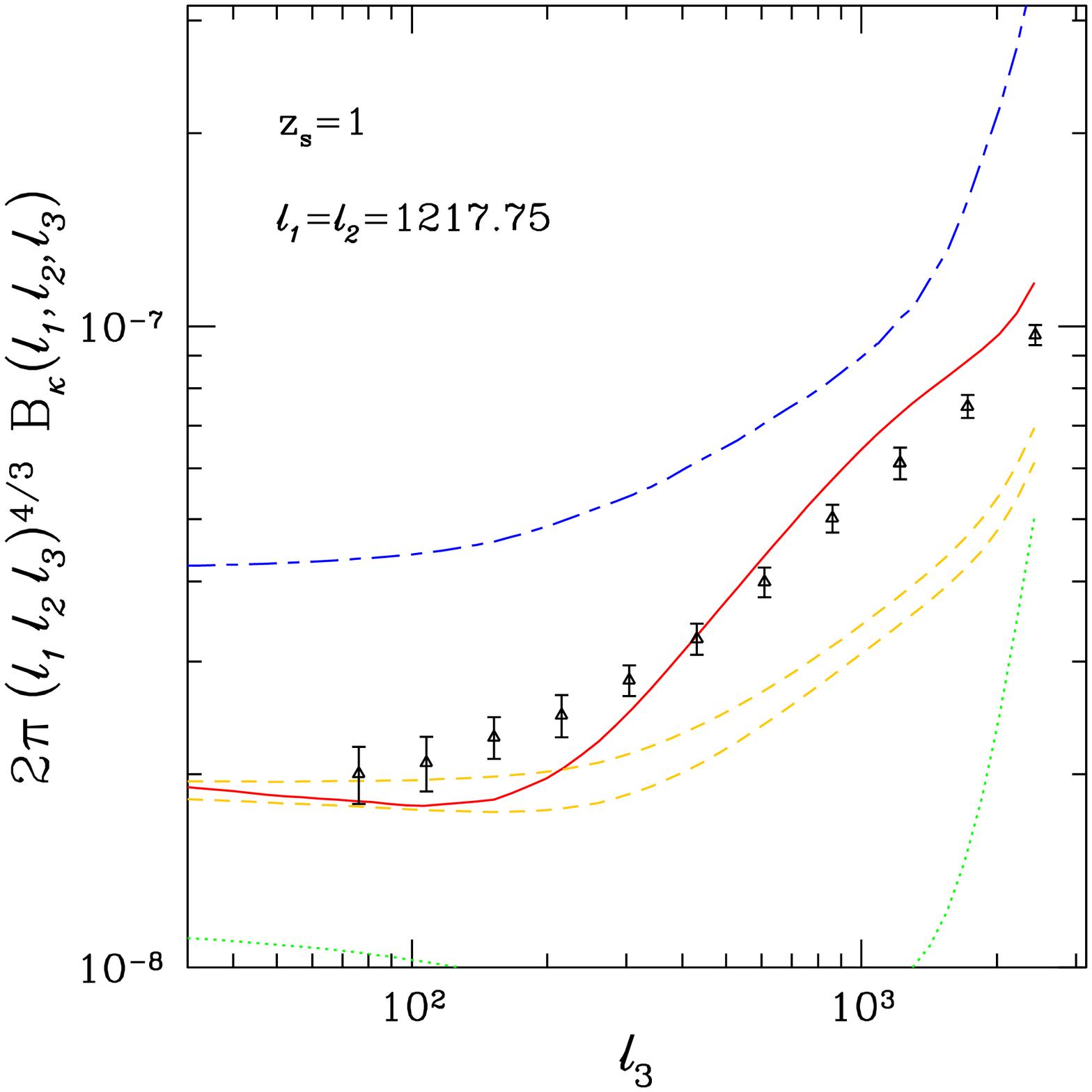}}
\epsfxsize=6.05 cm \epsfysize=5.4 cm {\epsfbox{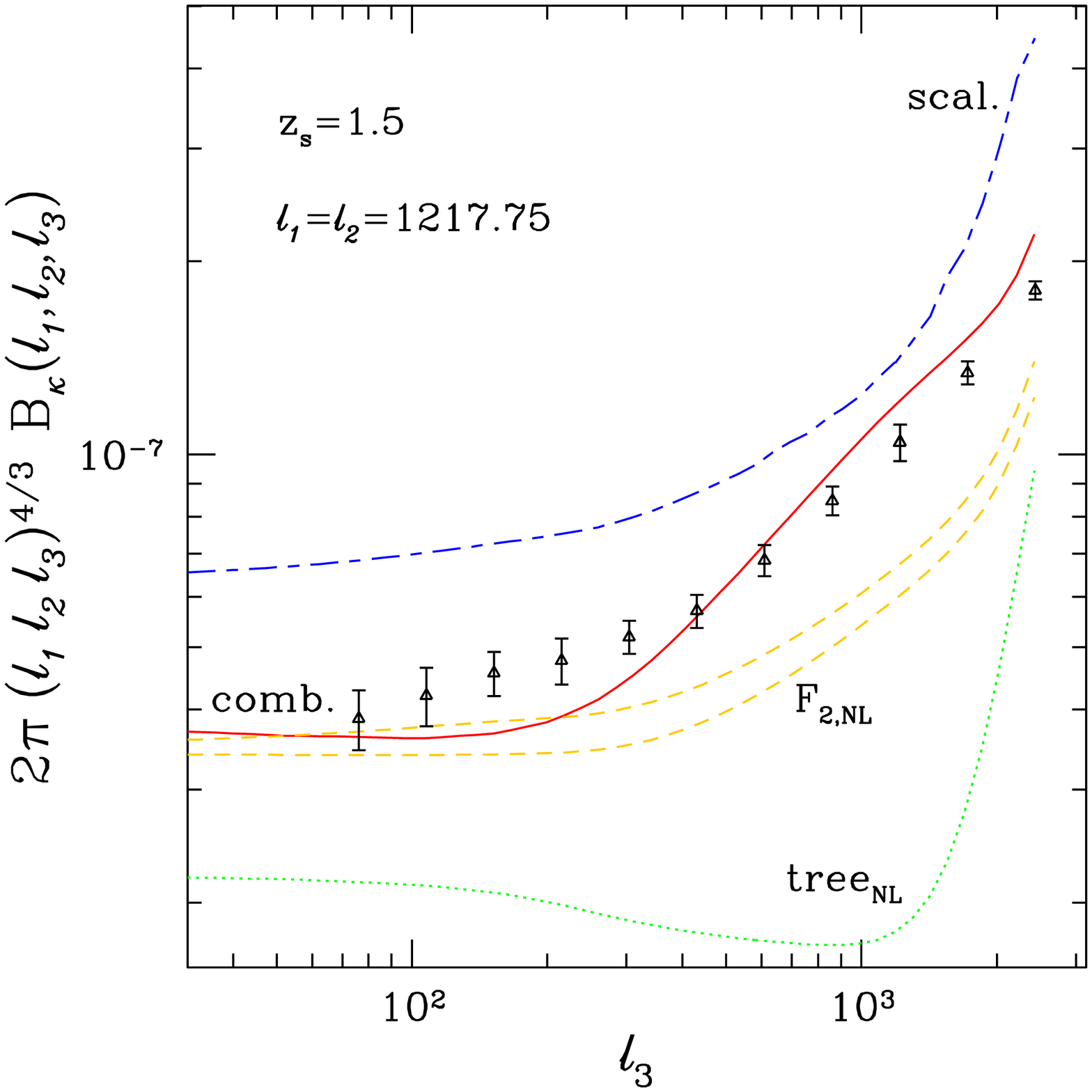}}
\end{center}
\caption{Convergence bispectrum $\Bkappa(\ell_1,\ell_2,\ell_3)$ for isosceles 
configurations, where $\ell_1=\ell_2$. The first two rows show the dependence on the
common side $\ell_1=\ell_2$, at fixed $\ell_3$ (in the weakly nonlinear regime
in the first row, and in the highly nonlinear regime in the second row), while the third
row shows the dependence on the third side $\ell_3$ at fixed $\ell_1=\ell_2$
(around the transition scales).
The symbols are the same as in Fig.~\ref{fig_Beq}.}
\label{fig_B_l3_l3_l1l2}
\end{figure*}

We first show in Fig.~\ref{fig_Beq} our results for the bispectrum for equilateral configurations. By construction, all curves must converge to the lowest-order
perturbative prediction (\ref{B-treeL-def}) on large scales. We can check that this is
indeed the case. As for the lensing power spectrum our simulations are
too small to reach this ``tree-order'' regime but they are consistent with this
asymptotic behavior.

Again, we recover the well-known fact that on the scales of interest lowest-order
perturbation theory is not sufficient to predict weak-lensing statistics and we must
take into account higher-order terms or nonperturbative contributions.
Using the nonlinear power within the ``tree-level'' expression, as in
Eq.(\ref{B-treeNL-def}), significantly increases the convergence bispectrum  
and improves the overall shape but it remains far from the simulation results and
cannot be used for practical purposes.
Adding a modification to the kernel $F_2$, as in Eq.(\ref{B-F2NL-def}) using the
fitting formula from \cite{Scoccimarro2001a}, further improves the predictions
and it provides a good match to the simulations on weakly nonlinear scales.
However, this simple fitting procedure underestimates 
the power at high $\ell$.
This is partly related to the underestimate of the power spectrum on small scales
by the ``halo-fit'' fitting formula of \cite{Smith2003}.
Indeed, we can see that using in Eq.(\ref{B-F2NL-def}) the nonlinear 3D power 
spectrum provided by our model, which provides a good match to simulations for
both the 3D power \citep{Valageas2011e} and the 2D convergence power spectrum 
(Fig.~\ref{fig_Pl}) increases the power at high $\ell$. This improves the agreement
with the simulations but there remains a significant discrepancy.

The scale transformation (\ref{B-Pan-def}) fares better on weakly nonlinear scales
(at least for this equilateral configuration) but it greatly overestimates the power at
high $\ell$. The same breakdown was already observed for the 3D bispectrum
\citep{Pan2007,Valageas2011e}.
The best agreement with the numerical simulations is provided by our model
(\ref{Bk-halo})-(\ref{B-1H}). In particular, it is interesting to note the good
match on the transition scales, $\ell \sim 10^3$, which a priori are the most difficult
to reproduce since they are at the limit of validity of both perturbative approaches
(which break down at shell crossing) and halo models (which assume virialized halos).
A similarly good agreement was also observed for the 3D bispectrum in
\cite{Valageas2011e}.

At high $\ell$ we again predict more power than is measured in the simulations, but
as for the lensing power spectrum this is at least partly due to the lack of
small-scale power in the simulations because of the finite resolution.
Thus, we again plot in Fig.~\ref{fig_Beq} the vertical arrows that were plotted
in Fig.~\ref{fig_Pl}. Since the bispectrum typically scales as the square of the
power spectrum this should roughly correspond to an accuracy threshold of
about $10\%$ for the simulations. We can check that our model agrees with the
numerical results up to this multipole.

\subsection{Isosceles triangles}
\label{Iso}

We now plot in Fig.~\ref{fig_B_l3_l3_l1l2} the convergence bispectrum for isosceles
configurations, where $\ell_1=\ell_2$. Instead of the angle $\vartheta_{12}$ between
the two sides $\ell_1=\ell_2$ we show the dependence on the length of either the
two equal sides or of the third side. This avoids putting all ``squeezed'' triangles in a
narrow region around $\vartheta_{12}= 0$. 

As for the equilateral configurations shown in Fig.~\ref{fig_Beq}, we can check
that the ``tree-order'' bispectrum (\ref{B-treeL-def}) is not the dominant contribution
on these scales, inserting the nonlinear power spectrum as in (\ref{B-treeNL-def})
is not sufficient, and the scale transformation (\ref{B-Pan-def}) breaks down too
early on nonlinear scales. Thus, the only two reasonable models are the
expression (\ref{B-F2NL-def}), which involves fitting formulas for both the
nonlinear power spectrum and the kernel $F_{2,\rm NL}$, and our model
(\ref{Bk-halo})-(\ref{B-1H}).

In the first row, where $\ell_3\simeq 107$ (the analysis uses the flat-sky approximation
hence $\ell$ is not necessarily an integer), both curves (\ref{B-F2NL-def}) and
(\ref{Bk-halo}) agree rather well with the numerical simulations, except that for
extremely ``squeezed'' configurations, $\ell_1 \ga 100 \ell_3$, our model yields
a downturn which underestimates the bispectrum whereas the fitting model
(\ref{B-F2NL-def}) remains in good agreement with the numerical data.
We shall come back to this point in Sect.~\ref{Iso-1H} below, where we shall find out
that this is probably due to an inaccuracy of the 2-halo contribution $B_{\rm 2H}$.
On the other hand, using the nonlinear power spectrum $P_{\rm tang}(k)$ in
Eq.(\ref{B-F2NL-def}) increases the power on small scales, as in Fig.~\ref{fig_Beq},
which spoils the match to simulations for these extremely ``squeezed'' configurations.

In the second row, which is farther into the nonlinear regime, we find in agreement
with the equilateral case of Fig.~\ref{fig_Beq} that the fitting formula (\ref{B-F2NL-def})
underestimates the bispectrum whereas our model provides a reasonable match to the
numerical simulations. More importantly, Eq.(\ref{B-F2NL-def}) does not reproduce
the shape of the bispectrum as a function of $\ell_1$, since it does not capture the
falloff that appears for $\ell_1 \ga 10 \ell_3$, and this is not cured by using a more
accurate 3D power spectrum. In contrast, our model (\ref{Bk-halo}) correctly
predicts this shape as well as the overall amplitude.

In the third row, which corresponds to transition scales in the mildly nonlinear
regime, neither model shows a perfect match to the simulations but our prediction
fares significantly better and still provides a reasonable agreement.

These results show that it is difficult to reproduce the bispectrum for a wide variety
of configurations and scales by using a global ansatz such as (\ref{B-F2NL-def}).
Approaches such as the one studied in this paper, which are built on
explicit and physical ingredients, are better controlled and offer an easier route to
systematic improvement. Indeed, by improving the accuracy of each ingredient
(e.g., including higher orders of perturbation theory or more complex halo profiles)
one should reach a higher accuracy for the final 3D and 2D statistics for any configuration
and scale (although transition regimes may still prove difficult).
Moreover, by splitting the problem into several elements it is easier to make progress
by improving each contribution in turn.

\section{Relative importance of the different contributions}
\label{contributions}

\begin{figure*}
\begin{center}
\epsfxsize=6.1 cm \epsfysize=5.4 cm {\epsfbox{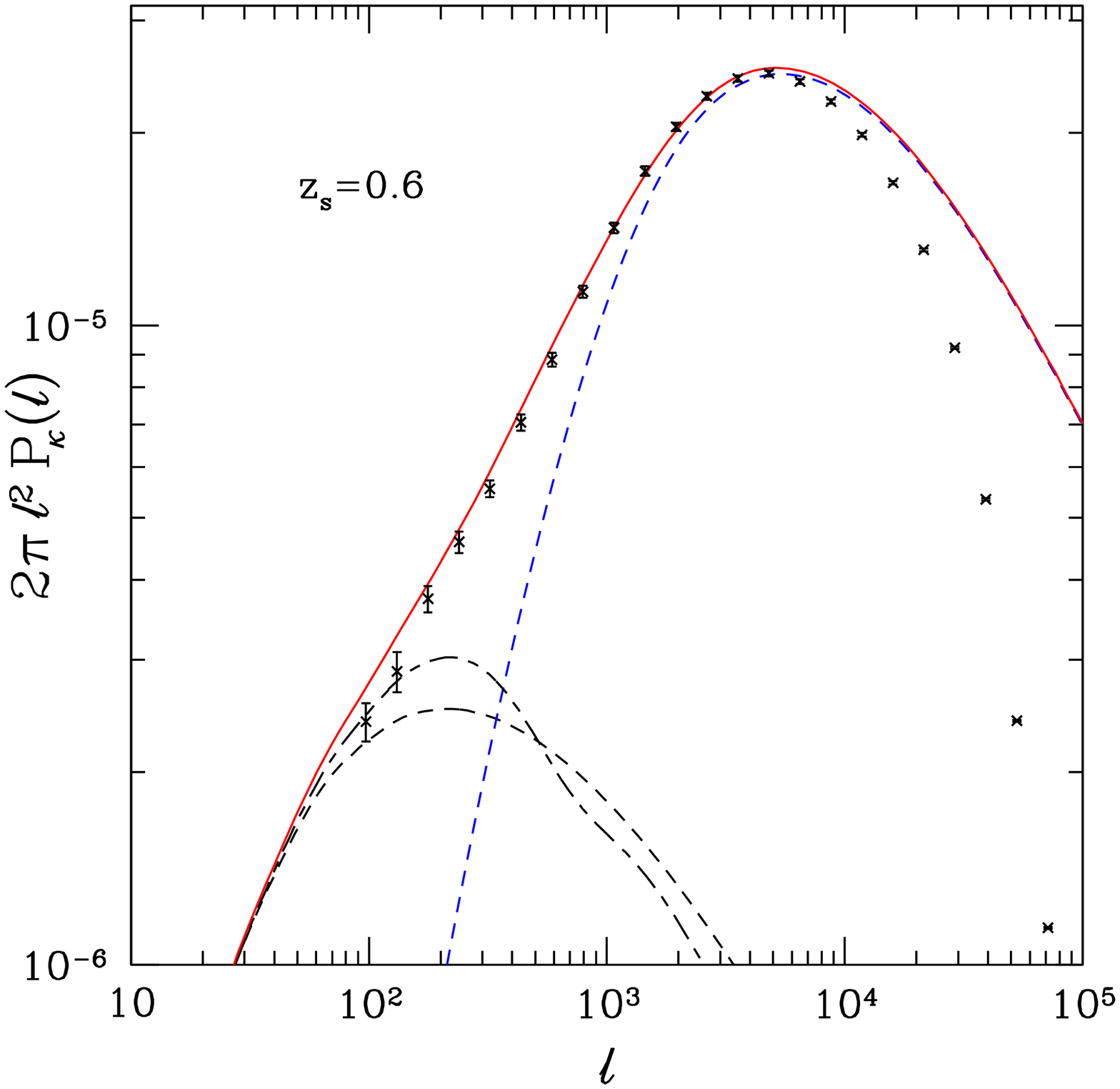}}
\epsfxsize=6.05 cm \epsfysize=5.4 cm {\epsfbox{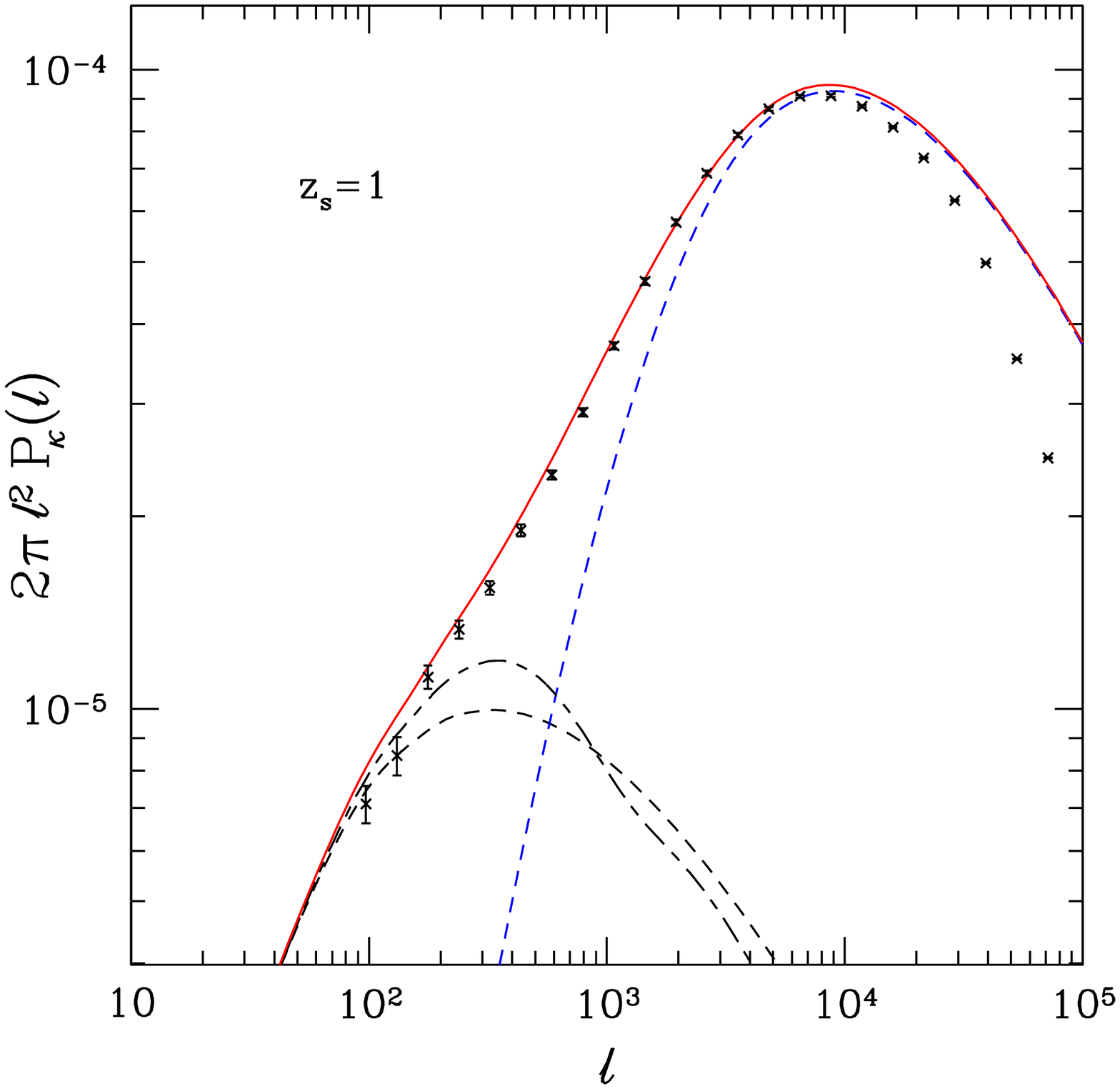}}
\epsfxsize=6.05 cm \epsfysize=5.4 cm {\epsfbox{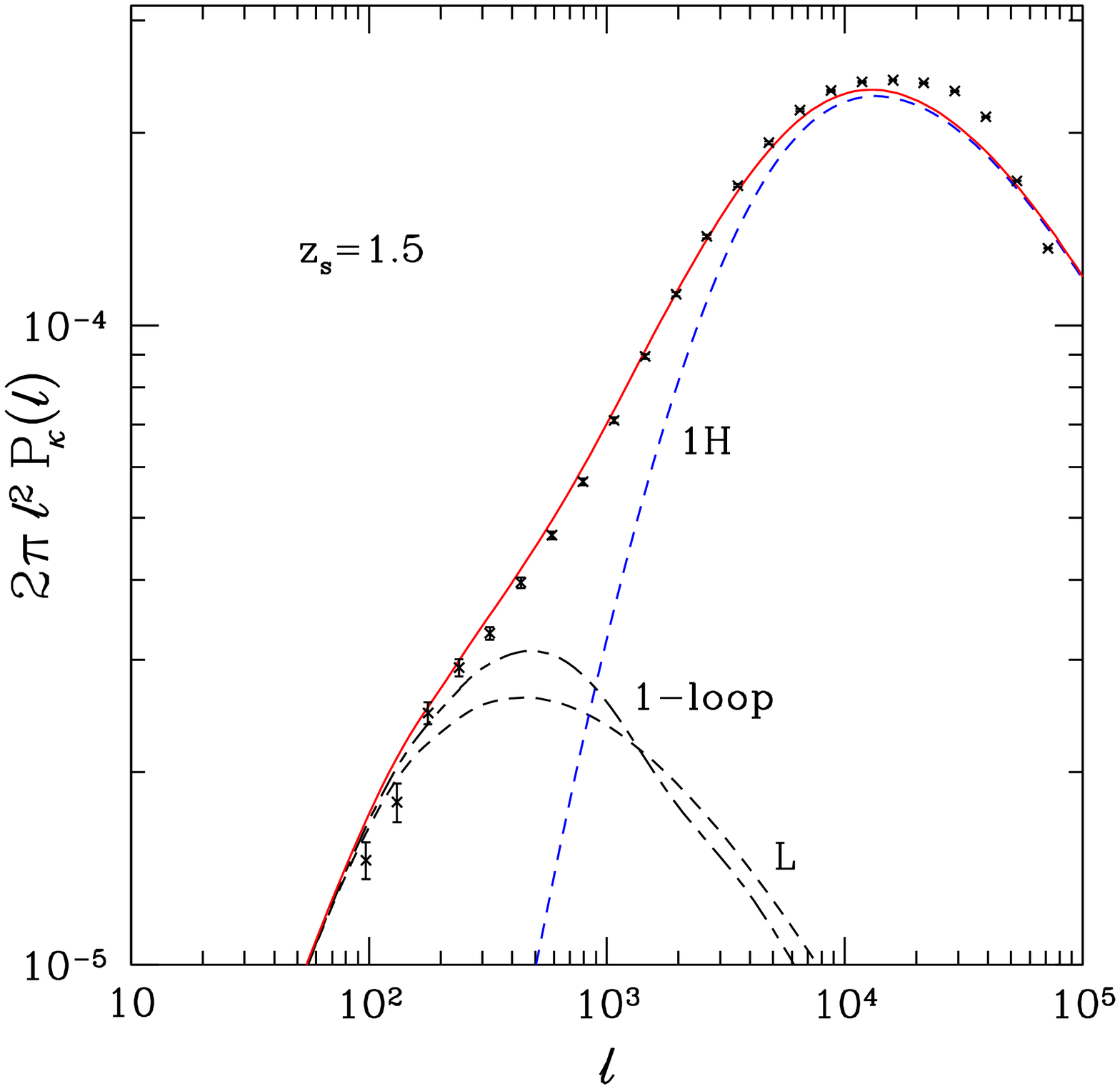}}
\end{center}
\caption{Convergence power spectrum for sources at redshifts $z_s=0.6, 1$, and
$1.5$. The points are the results from numerical simulations with $3-\sigma$
error bars.
The low black dashed line ``L'' is the linear power, the middle black dot-dashed line
``1-loop'' is the result from the 2-halo contribution (\ref{Pk-2H}), where we use a perturbative 
resummation that is complete up to 1-loop order, the upper blue dashed line ``1H'' is
the result from the 1-halo contribution (\ref{Pk-2H}), and the red solid
line is the result of our model (\ref{P-tang-1})-(\ref{P-tang-2}).}
\label{fig_Pl_1H}
\end{figure*}

\begin{figure*}
\begin{center}
\epsfxsize=6.1 cm \epsfysize=5.4 cm {\epsfbox{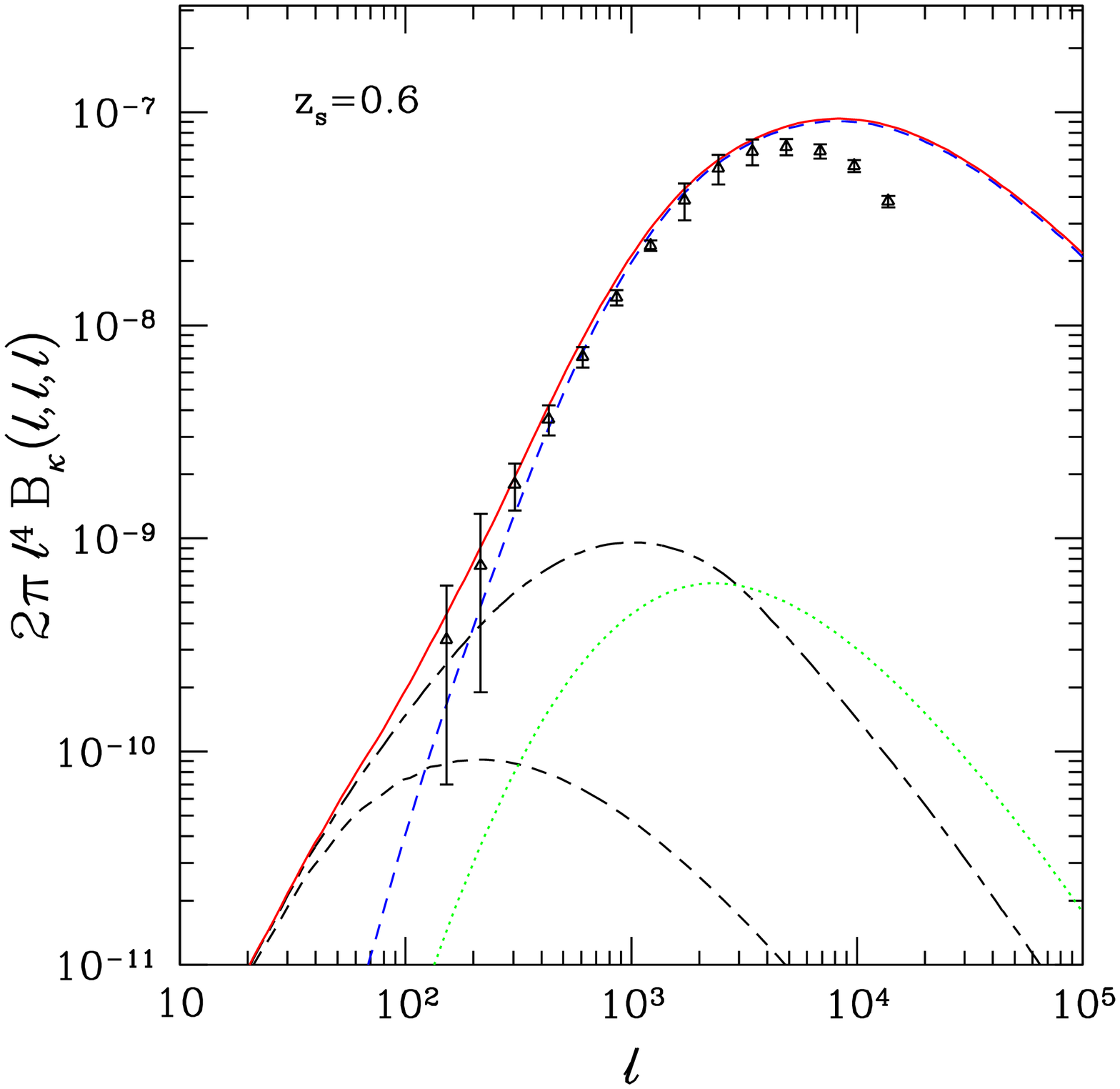}}
\epsfxsize=6.05 cm \epsfysize=5.4 cm {\epsfbox{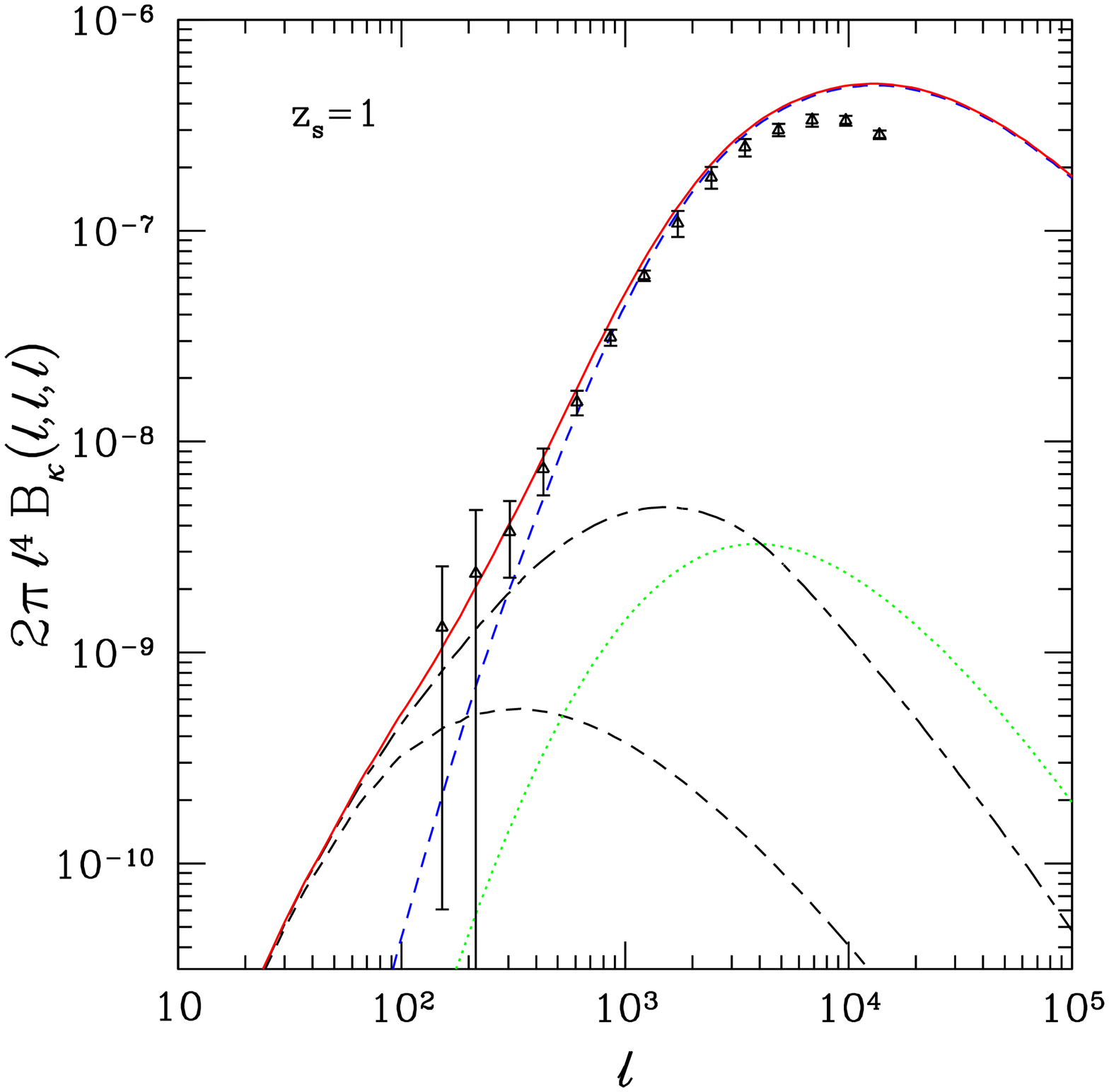}}
\epsfxsize=6.05 cm \epsfysize=5.4 cm {\epsfbox{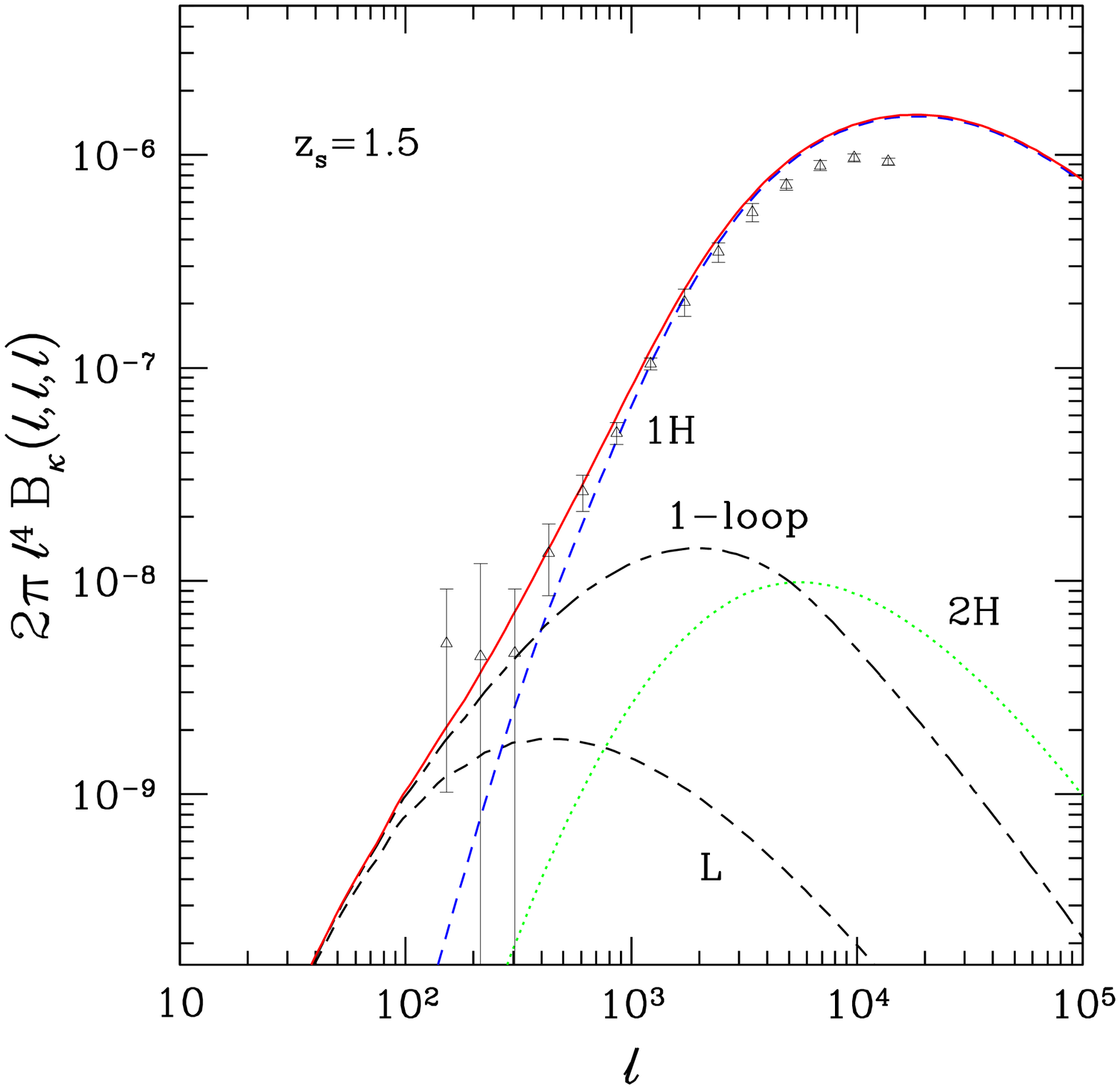}}
\end{center}
\caption{Convergence bispectrum for sources at redshifts $z_s=0.6, 1$, and
$1.5$, for equilateral triangles.
The points are the results from numerical simulations with $3-\sigma$ error bars.
The low black dashed line ``L'' is the tree-order bispectrum (\ref{B-treeL-def}),
the ``1-loop'' dot-dashed black line is the prediction of 1-loop standard perturbation
theory, which is identified with our 3-halo term (\ref{B-3H}), the green dotted line
``2H'' is the 2-halo contribution (\ref{B-2H}) and the upper blue dashed line
``1H'' is the 1-halo contribution (\ref{B-1H}).}
\label{fig_Beq_1H}
\end{figure*}

We have seen in the previous section that our model, based on a combination
of perturbation theories and halo models, provides a good match to numerical
simulations. This means that such an approach can be useful to obtain
predictions for weak-lensing statistics for a variety of cosmologies, which is   
an important goal for observational and practical purposes.

A second advantage of our approach, as compared with simple fitting formulas
(or direct simulations), is that we can easily evaluate and compare the different 
contributions that eventually add up to the signal that can be measured in
weak-lensing surveys. Thus, we can distinguish the various perturbative terms
as well as the nonperturbative contributions associated with 1-halo or 2-halo
terms.

This is useful to estimate the accuracy that can be aimed at in weak-lensing statistics,
as a function of scales, since different contributions suffer from different theoretical
uncertainties. For instance, the perturbative contributions are obtained in a systematic
and rigorous manner for any cosmology, and the accuracy is only limited by the
order of the perturbative expansion, whereas ``halo contributions'', which depend
on the halo profiles and mass functions, are phenomenological ingredients that
involve at some stage input from numerical simulations and do not offer systematic
 routes to arbitrarily high accuracy.

This also allows us to understand which aspects of the dynamics are probed
by weak-lensing statistics, so that depending on one's goals (e.g., to constrain
cosmological parameters or to estimate halo properties themselves) one can
focus on the appropriate range of the weak-lensing signals.

\subsection{Convergence power spectrum}
\label{Convergence-power-spectrum-1H}

We plot our results for the convergence power spectrum in Fig.~\ref{fig_Pl_1H},
showing the underlying 2-halo and 1-halo contributions in addition to the full model
curve that was already shown in Fig.~\ref{fig_Pl}.
We can see that including the 1-loop perturbative term yields slightly more
power on quasi-linear scales and helps to obtain a good match to the simulations.
At high $\ell$ the 2-halo contribution goes back close to the linear power thanks to
the partial resummation of higher orders. As recalled in Sect.~\ref{3D-power},
this is a useful improvement over the standard 1-loop perturbation theory because
it ensures that at high $k$ for the 3D power, and at high $\ell$ for the lensing power,
the 2-halo term does not give significant contributions.
At high $\ell$ around its peak, the lensing power becomes dominated by the
1-halo term but as for the 3D power there remains a wide intermediate range, about a
decade over $\ell$. In our model, the 3D power on these transition scales is described
by the interpolation (\ref{P-tang-2}), which was seen in \cite{Valageas2011e}
to provide a good match to numerical simulations. 
As expected, Fig.~\ref{fig_Pl_1H} shows that this also provides a good interpolation
for the 2D convergence power spectrum.
It may be possible to reduce this range on its low-$\ell$ side by including higher 
orders of perturbation theory in the 2-halo term, but it is likely that the high-$\ell$
side receives significant contributions from nonperturbative terms, associated with
shell crossing \citep{Valageas2011a}, that are not well described either by a simple
1-halo term (e.g., associated with dense filaments or infalling regions close to the
virial radius).
In any case, Fig.~\ref{fig_Pl_1H} clearly shows how the convergence power spectrum
depends on large-scale perturbative density fluctuations or on small scale halo
properties, as the multipole $\ell$ varies.

Future surveys plan to measure the weak lensing power spectrum on these scales,
$50 < \ell < 7000$, with an accuracy of about $1\%$ for LSST \citep{LSSTScienceCollaborations2009}
or even slightly better for Euclid \citep{Laureijs2011}.
Our numerical simulations do not allow us to check our model down to this accuracy.
However, the comparison between our predictions and the ``halo-fit'' in Fig.~\ref{fig_Pl}
already shows that this is beyond the accuracy of current models and is a very challenging
theoretical goal.
Our model already provides an accuracy of about $1\%$ on perturbative scales
and $10\%$ on the transition and highly nonlinear scales
(this is the typical accuracy checked on 3D simulations in \citet{Valageas2011e}).
Figure~\ref{fig_Pl_1H} explicitly shows that for $100<\ell<1000$ the one-loop
correction to the linear power is required to reach this accuracy. Unfortunately, for $\ell > 200$
this is not sufficient since higher-order terms or non-perturbative corrections cannot be 
neglected and the power spectrum is sensitive to the transition scales, described by the 
interpolation (\ref{P-tang-2}). This suggests that for $\ell > 500$ a percent accuracy
is probably beyond the reach of systematic perturbative approaches, which cannot go
beyond shell crossing \citep{Valageas2011a}, and requires dedicated numerical simulations.

\subsection{Convergence bispectrum}
\label{Convergence-bispectrum-1H}

We now study the contributions from various terms to the lensing bispectrum.
From Eq.(\ref{Bk-halo}) we now have
three contributions, associated with the 3-halo, 2-halo and 1-halo terms,
the 3-halo contribution being identified with the perturbative contribution.

\subsubsection{Equilateral triangles}
\label{Equilateral-1H}

\begin{figure*}
\begin{center}
\epsfxsize=6.1 cm \epsfysize=5.4 cm {\epsfbox{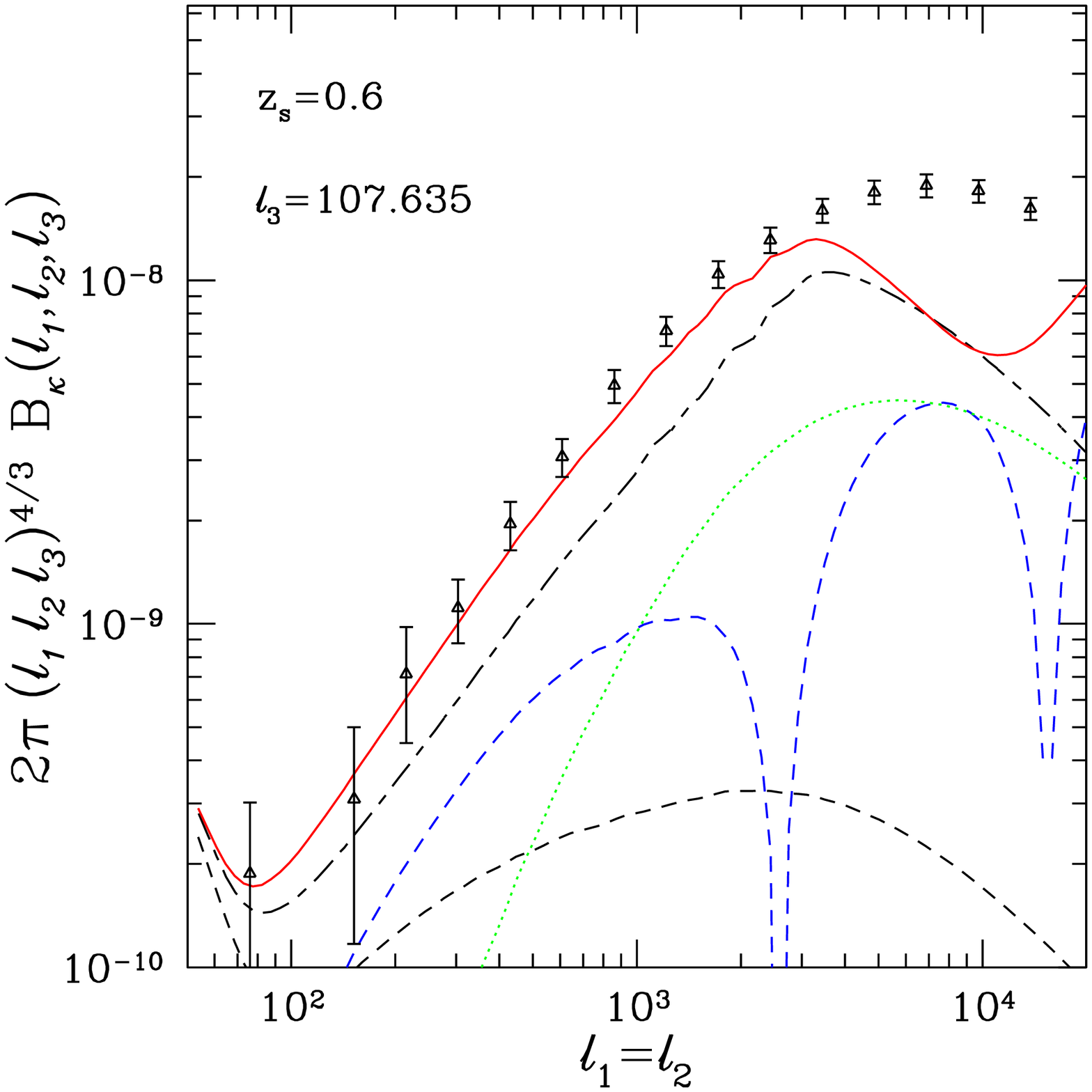}}
\epsfxsize=6.05 cm \epsfysize=5.4 cm {\epsfbox{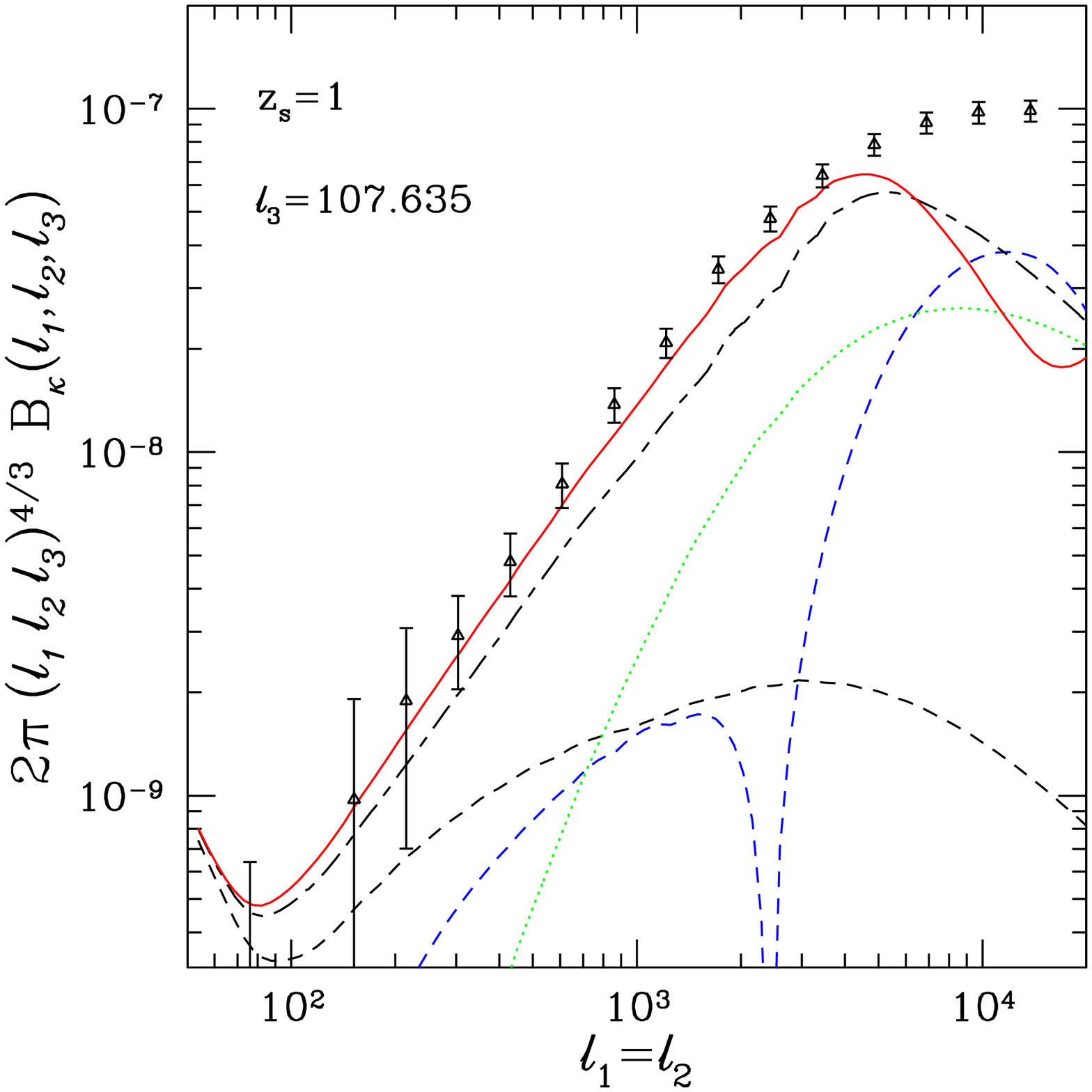}}
\epsfxsize=6.05 cm \epsfysize=5.4 cm {\epsfbox{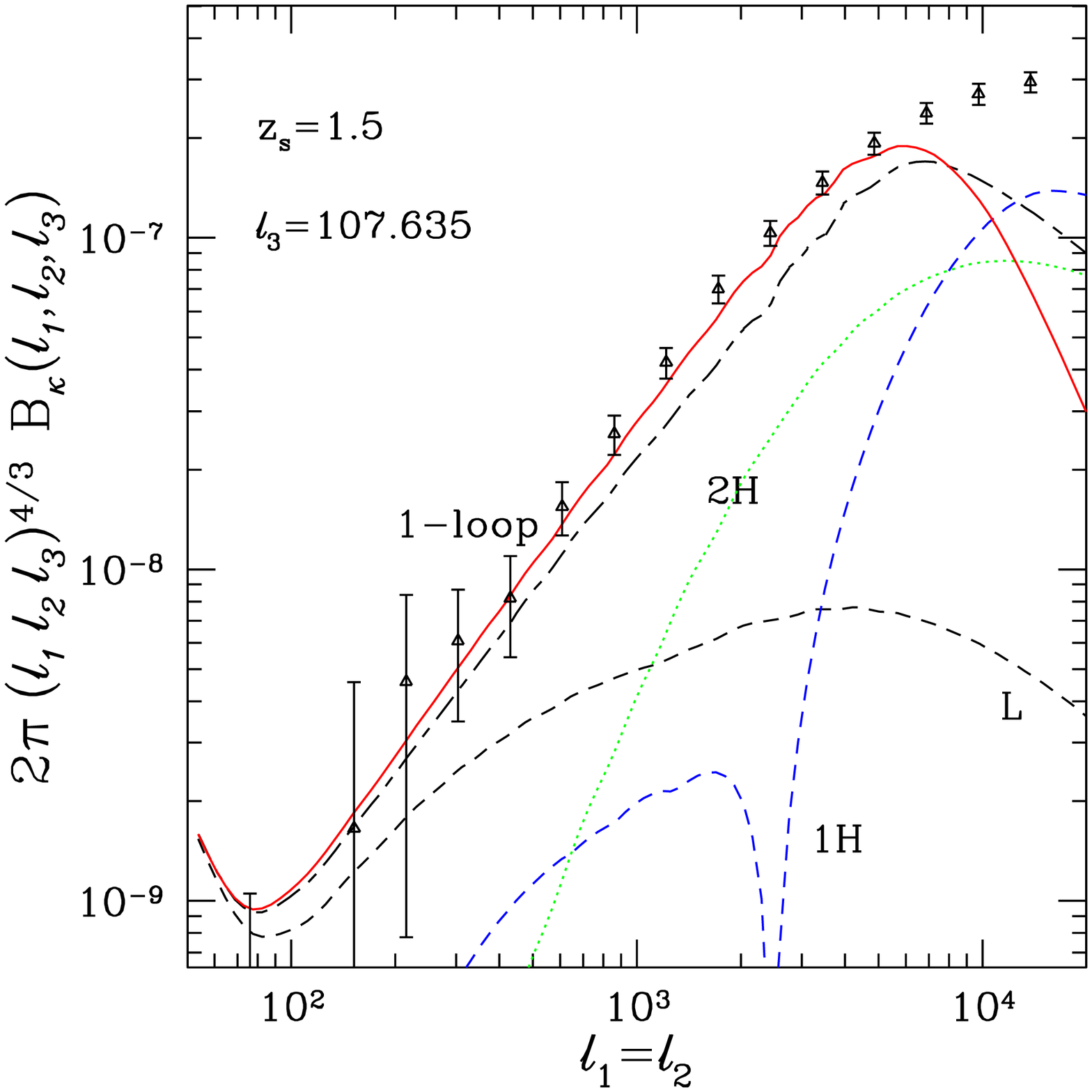}}\\
\epsfxsize=6.1 cm \epsfysize=5.4 cm {\epsfbox{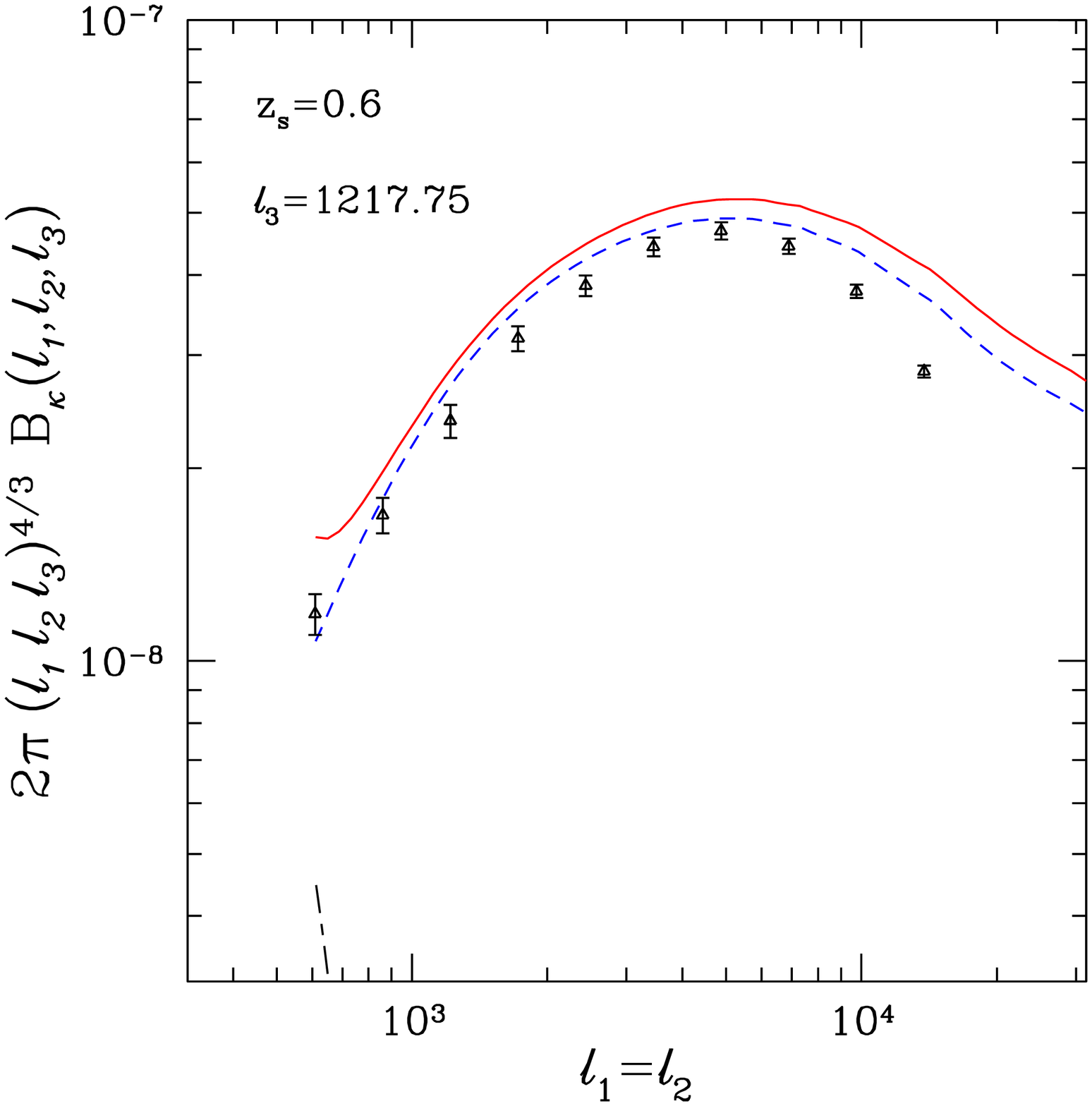}}
\epsfxsize=6.05 cm \epsfysize=5.4 cm {\epsfbox{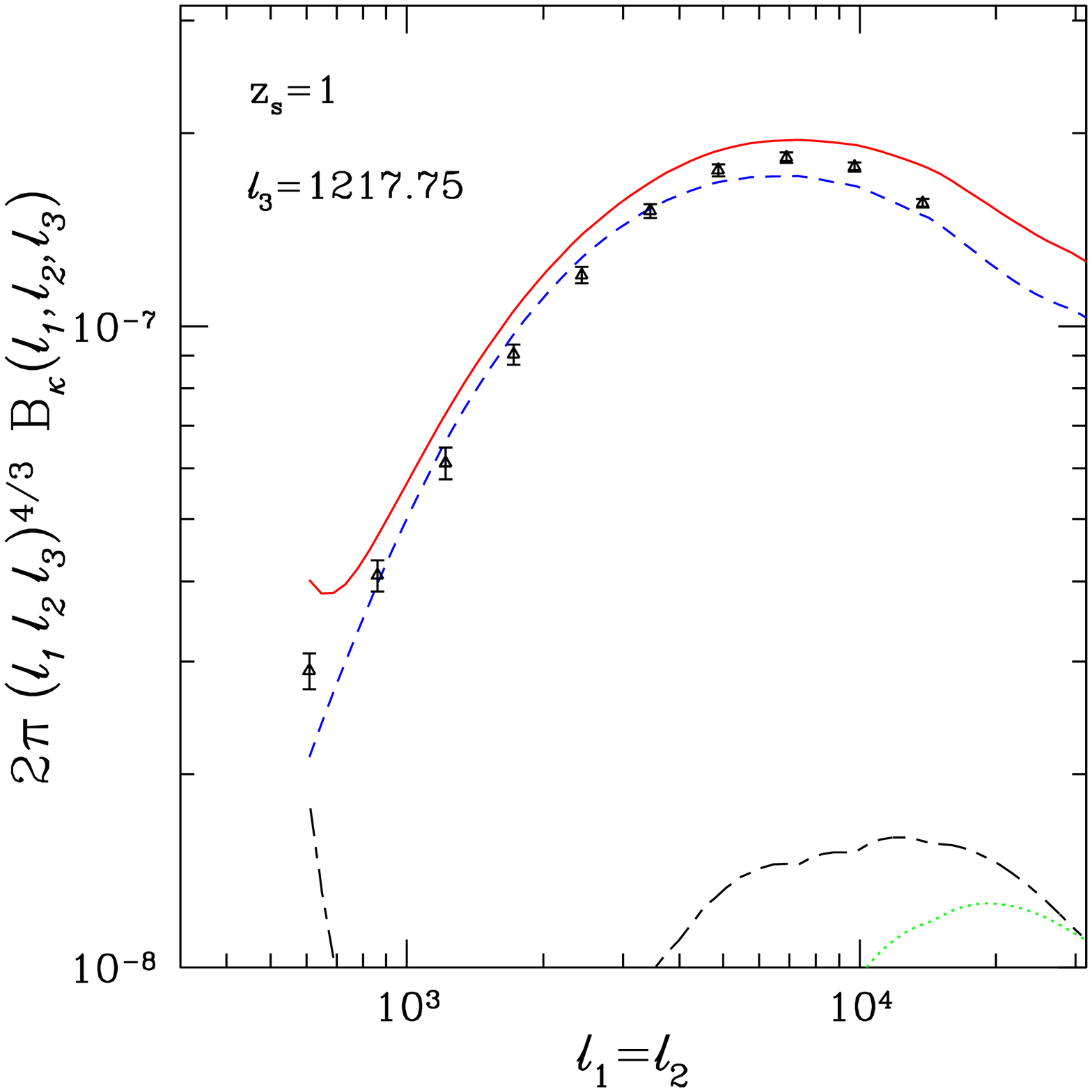}}
\epsfxsize=6.05 cm \epsfysize=5.4 cm {\epsfbox{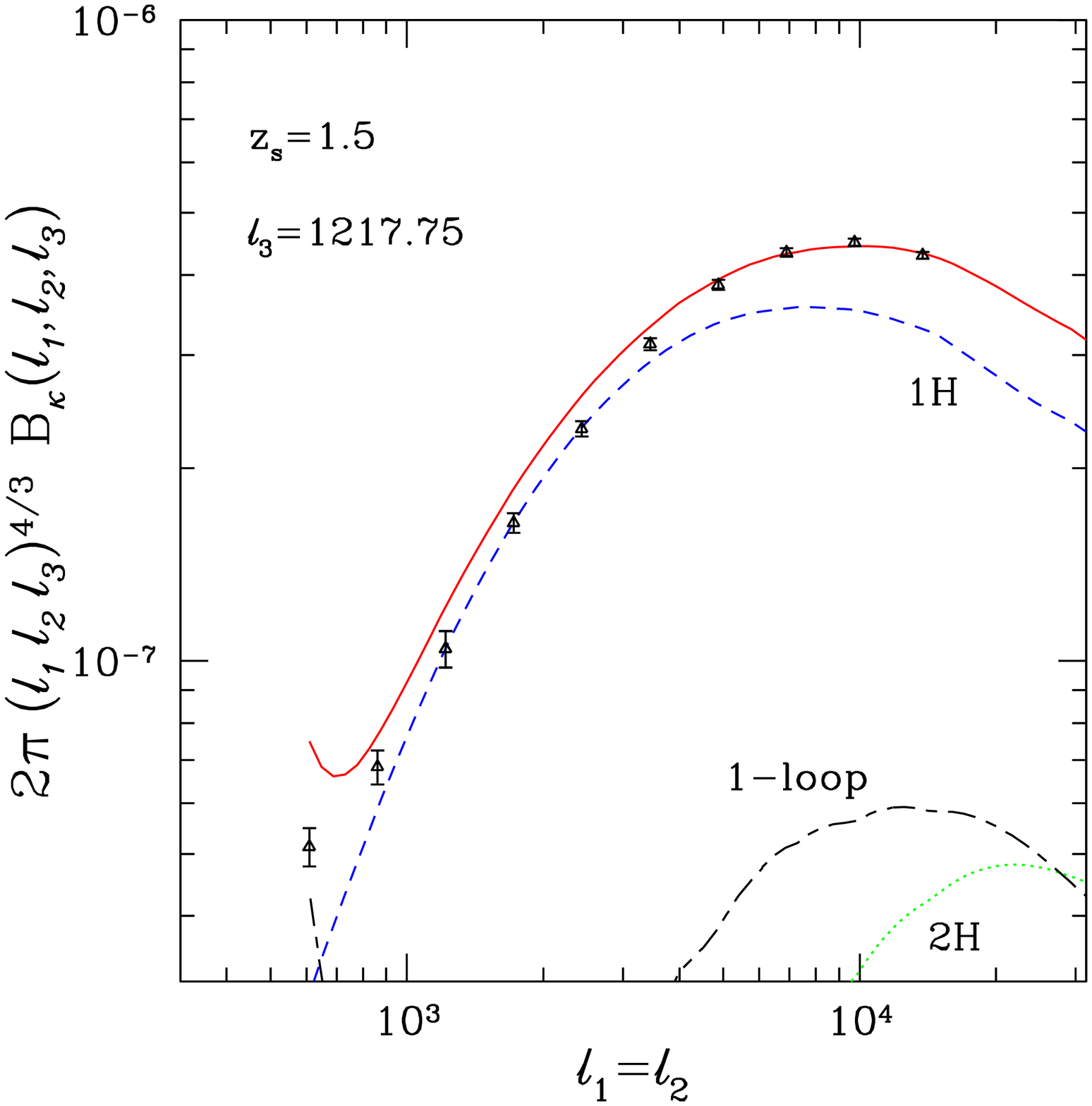}}\\
\epsfxsize=6.1 cm \epsfysize=5.4 cm {\epsfbox{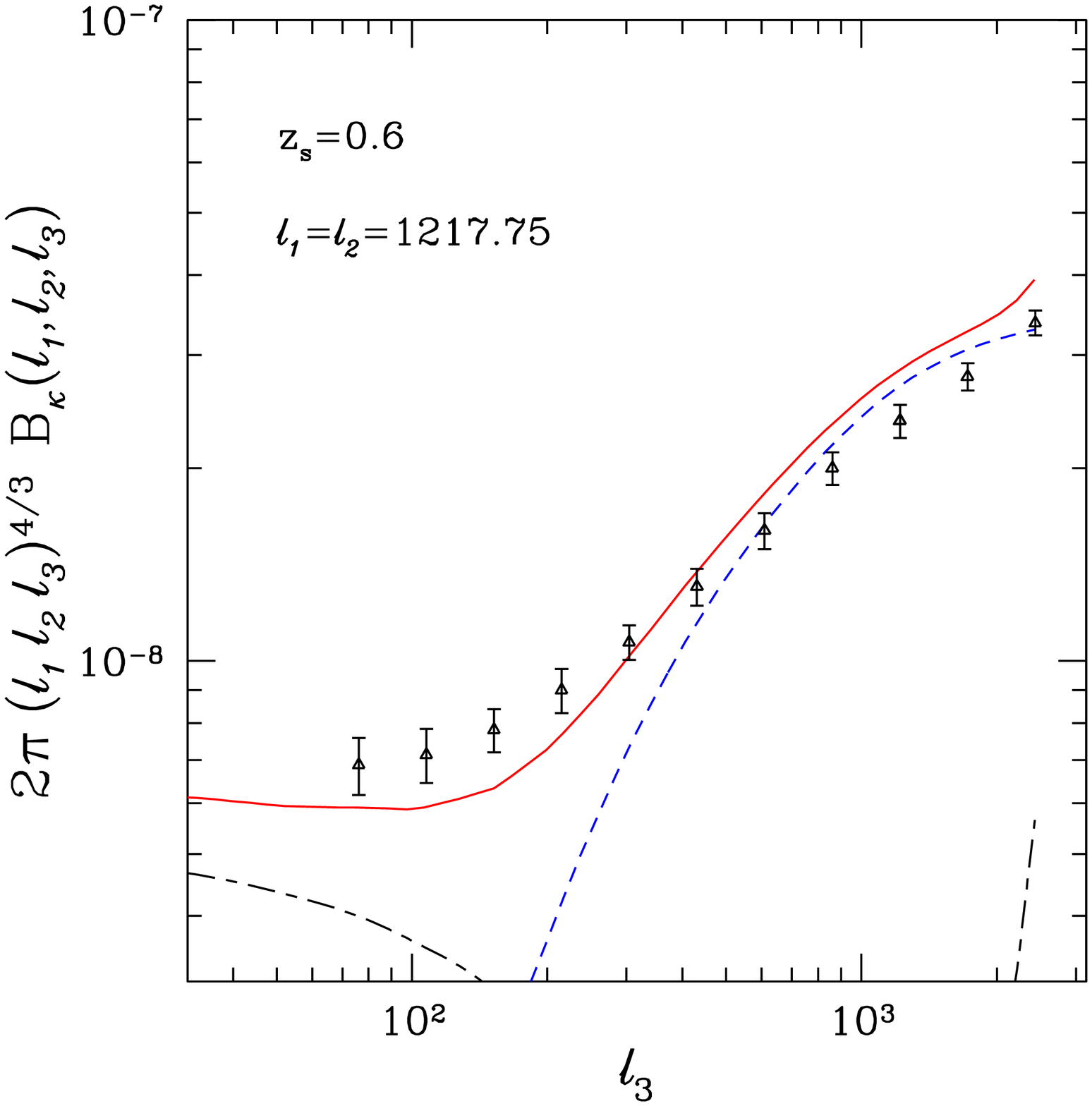}}
\epsfxsize=6.05 cm \epsfysize=5.4 cm {\epsfbox{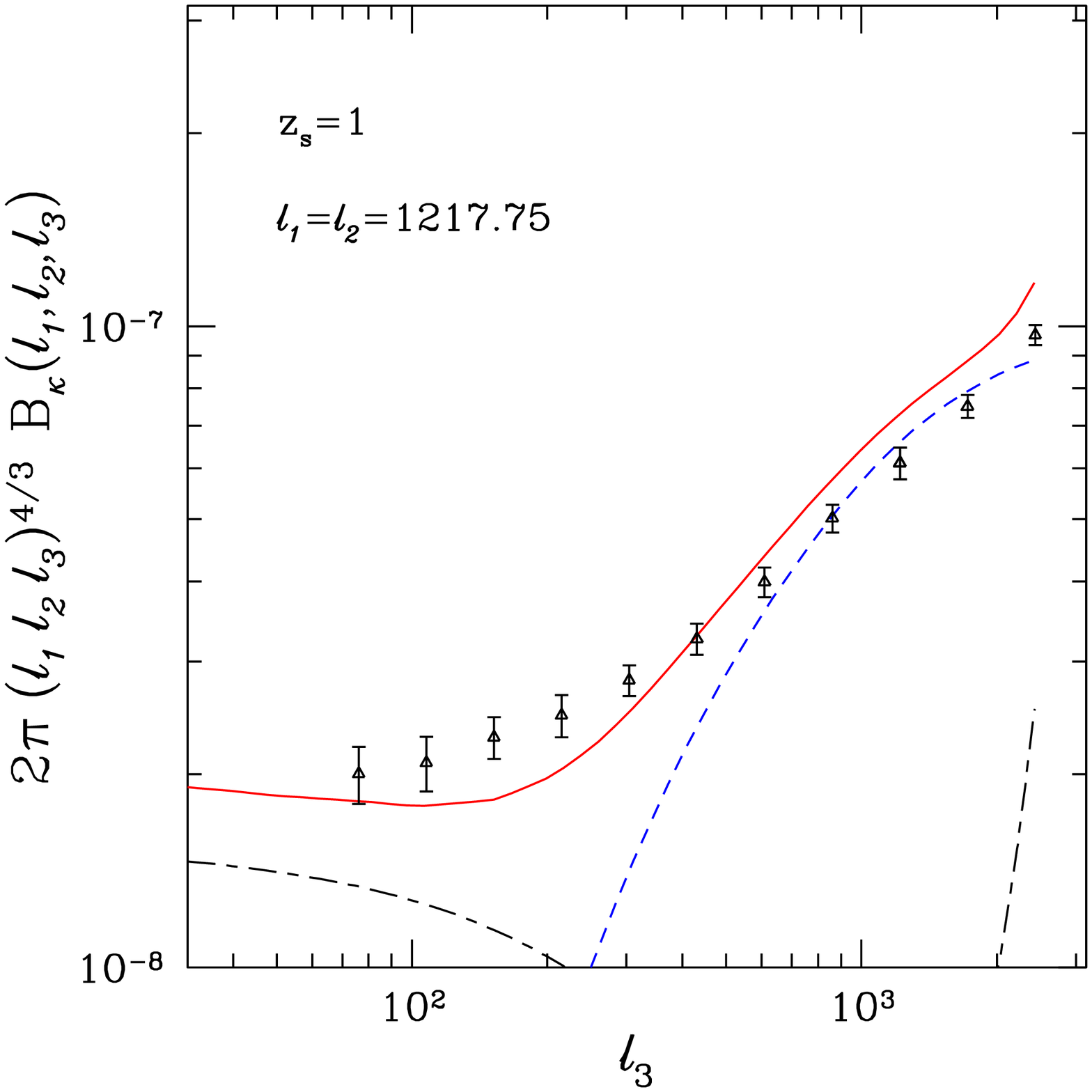}}
\epsfxsize=6.05 cm \epsfysize=5.4 cm {\epsfbox{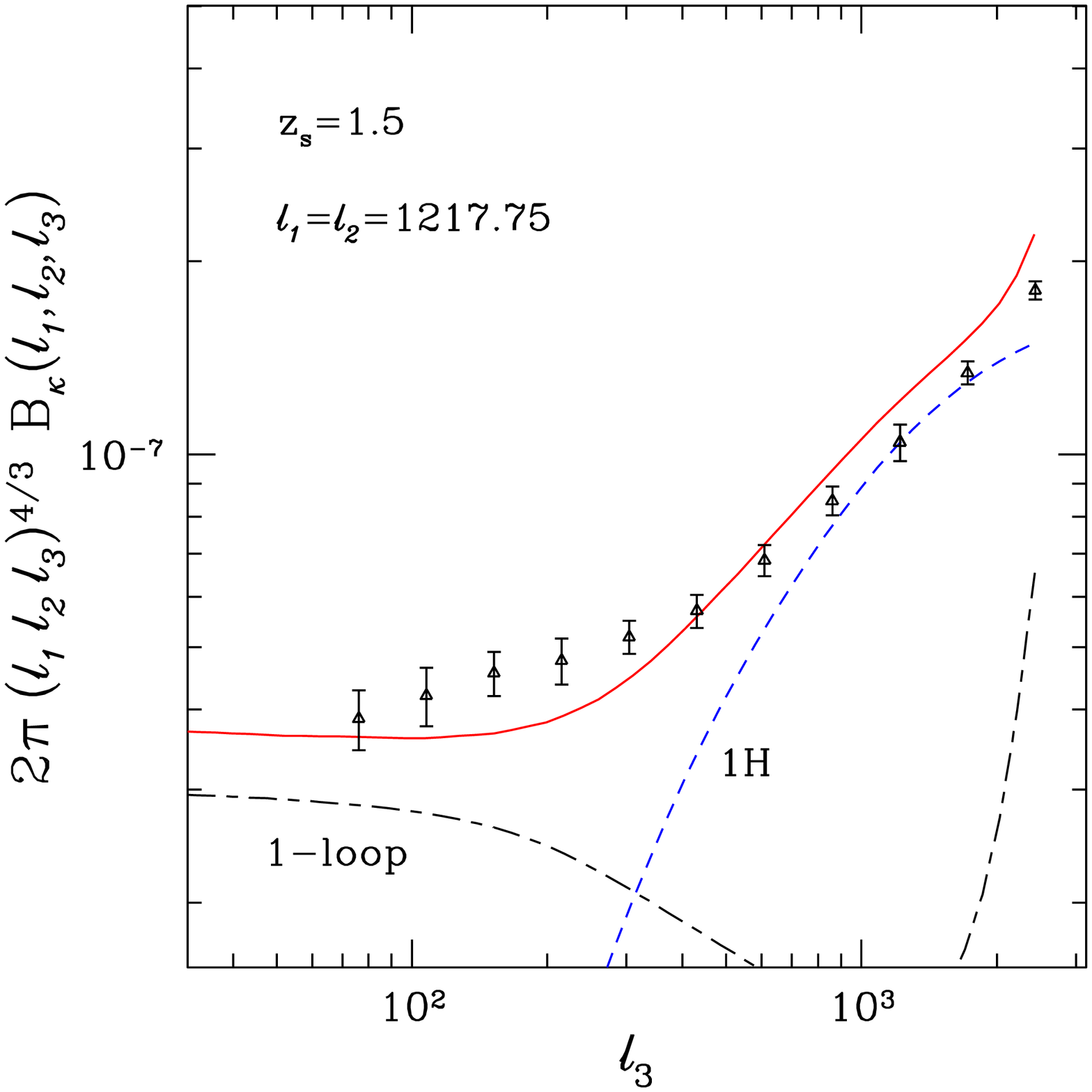}}
\end{center}
\caption{Convergence bispectrum $\Bkappa(\ell_1,\ell_2,\ell_3)$ for isosceles 
configurations, as in Fig.~\ref{fig_B_l3_l3_l1l2}.
The symbols are the same as in Fig.~\ref{fig_Beq_1H}. In the first row we show
the absolute value of the 1-halo contribution, which becomes negative at
$\ell_1 \sim 2500$.}
\label{fig_B_l3_l3_l1l2_1H}
\end{figure*}

We plot our results for the convergence bispectrum for equilateral configurations
in Fig.~\ref{fig_Beq_1H}, showing the various contributions in addition to the full
model curve that was already shown in Fig.~\ref{fig_Beq}.
As recalled in Sect.~\ref{3D-power}, the 3-halo term is identified with the perturbative
prediction and in this paper we use standard perturbation theory at one-loop order
for this contribution. As for the 3D bispectrum \citep{Valageas2011e}, and contrary
to the power spectrum, this gives a contribution that becomes
negligible on small scales, so that it is not necessary to use a resummation scheme
or to add a nonperturbative cutoff to ensure a good high-$\ell$ behavior.
On the other hand, contrary to the power spectrum we can see that going to 1-loop
order now makes a strong improvement over the tree-order result.
This feature was already noticed for the 3D density field \citep{Sefusatti2010,Valageas2011e}.
In fact, combining this 1-loop perturbative contribution with the 2-halo and 1-halo
terms is sufficient to obtain a good match to the simulations, from the quasilinear
to the highly nonlinear scales. This suggests that higher orders of perturbation theory
do not significantly contribute to the bispectrum and that we already have
a reasonably successful model. For these equilateral triangles the 2-halo
term is also subdominant on all scales (by a factor $\sim 10$ at least).
This is a nice property since being a mixed term, which involves both large-scale
halo correlations and internal halo structures, it may be more difficult to predict
than the 3-halo term (derived from systematic perturbation theories) and the 1-halo
term (that only depends on internal halo profiles and mass function).
These various features were also observed for the 3D bispectrum \citep{Valageas2011e},
from which they directly derive through the relation (\ref{Bkappa-B}).

Again, our numerical simulations do not allow us to reach the percent accuracy aimed at by
future surveys \citep{LSSTScienceCollaborations2009,Laureijs2011}, and Figs.~\ref{fig_Beq}
and \ref{fig_Beq_1H} show that this is a challenging goal. As noticed above, including
1-loop corrections is clearly important to reach this accuracy.
A nice feature is that without introducing any interpolation on transition scales we already
have a reasonably good model and that the 2-halo term contributes to about $1\%$
or less on most scales. Then, our model already provides an accuracy of $10\%$
(this is the typical accuracy checked on 3D simulations in \citet{Valageas2011e})
on the scales shown in Fig.~\ref{fig_Beq_1H}, and better at low $\ell$ in the
perturbative regime. At high $\ell$, $\ell > 10^4$, including the effects of baryons
on halo profiles (as compared with dark matter only N-body simulations) is probably
necessary to reach percent accuracy, but this is beyond the scope of this paper.

\subsubsection{Isosceles triangles}
\label{Iso-1H}

We turn to isosceles configurations in Fig.~\ref{fig_B_l3_l3_l1l2_1H}.
As for the equilateral case plotted in Fig.~\ref{fig_Beq_1H}, the upper row shows
that on weakly nonlinear scales the 1-loop contribution makes a significant improvement
over the tree-order prediction and allows us to obtain a good match to the numerical
simulations. 
Moreover, we can see that our underestimate of the bispectrum for very ``squeezed''
triangles, at $\ell_1 \ga 100 \ell_3$ with $\ell_3\simeq 107$, which was already
noticed in Fig.~\ref{fig_B_l3_l3_l1l2}, corresponds to a change of regime.
Indeed, this downturn follows the 1-loop prediction and it occurs at a point where the
2-halo and 1-halo contributions become important. Then, this discrepancy may be
understood as a difficulty of our simple model to reproduce the bispectrum in this
corner of configuration space because the transition from the 3-halo perturbative
regime to the 1-halo highly nonlinear regime is not sufficiently well described
(e.g., simple spherical halo models cannot be expected to describe very well
intermediate regions such as filaments and outer infalling regions of halos).
Alternatively, it may happen that adding higher orders of perturbation theory to the
3-halo contribution improves the agreement with simulations for these isosceles
triangles, without contributing much to the equilateral triangles where the agreement is
already rather good.
Another possibility is that our 2-halo term is not very accurate and gives too small
a contribution for these isosceles configurations.
This would be a natural and simple cure, because as seen in Fig.~\ref{fig_Beq_1H}
this 2-halo term is negligible for equilateral cases. Therefore, it should be possible
to ``tune'' the 2-halo term so as to match the isosceles simulation results of 
Fig.~\ref{fig_B_l3_l3_l1l2_1H} without damaging the good agreement obtained
for the equilateral case  in Fig.~\ref{fig_Beq_1H}.
This would also be a more natural explanation than a problem in the 1-halo term,
because the latter should be simpler to model (since by definition it only deals
with inner regions of halos) and it provides successful results in regimes where
it dominates the bispectrum, as shown by the second row in
Fig.~\ref{fig_B_l3_l3_l1l2_1H} and the high-$\ell$ part of Fig.~\ref{fig_Beq_1H}.

Since the problem found in the high-$\ell_1$ part of the upper row of 
Fig.~\ref{fig_B_l3_l3_l1l2_1H} corresponds to very ``squeezed'' triangles,
$\ell_1 \ga 100 \ell_3$, hence to a limited region of the configuration space for the
triplet $\{\vell_1,\vell_2,\vell_3\}$, and our goal is to study simple and systematic
models (i.e., that do not require fitting internal parameters for each new set of
cosmological parameters), we do not investigate in this paper how to improve
the model (e.g., through added complexity of the 2-halo term) to better describe
this regime. 
 
We can see in the second row, which is fully dominated by the 1-halo contribution,
that the curved shape of the bispectrum does not arise from a change of regime
and a switch from one contribution to the other as in the first row. This is a genuine
effect associated with the behavior of the 1-halo term that is well described by
our model. In contrast, as seen in Fig.~\ref{fig_B_l3_l3_l1l2}, this shape was not
well recovered by the ansatz (\ref{B-F2NL-def}).
This shows the benefits of our approach that explicitly combines perturbation
theory with a halo model, taking advantage of the good behavior of these two
ingredients in the regimes where they dominate.

In the third row, which corresponds to transition scales, we can see a change
of regime as the bispectrum is dominated by the 1-loop perturbative contribution
at low $\ell_3$ and by the 1-halo contribution at high $\ell_3$. In contrast to the
first row, the 2-halo contribution does not play a significant role and our model
provides a good agreement with the simulations on all scales.
This is a nice result since one could have expected transitions to be difficult to
reproduce. This again suggests that the mismatch found for extremely ``squeezed''
triangles in the first row may be due to the 2-halo term. Moreover, this shows that
a good modeling of the 3-halo (i.e. perturbative) and 1-halo terms is sufficient to
provide accurate predictions for the lensing convergence bispectrum for 
a wide variety of configurations including some transition regimes.
This agrees with the good agreement obtained for equilateral configurations,
which are also dominated by the 3-halo and 1-halo terms as seen in
Fig.~\ref{fig_Beq_1H}.

\section{Dependence on cosmology}
\label{Cosmology}

\begin{figure}
\begin{center}
\epsfxsize=8.6 cm \epsfysize=7.5 cm {\epsfbox{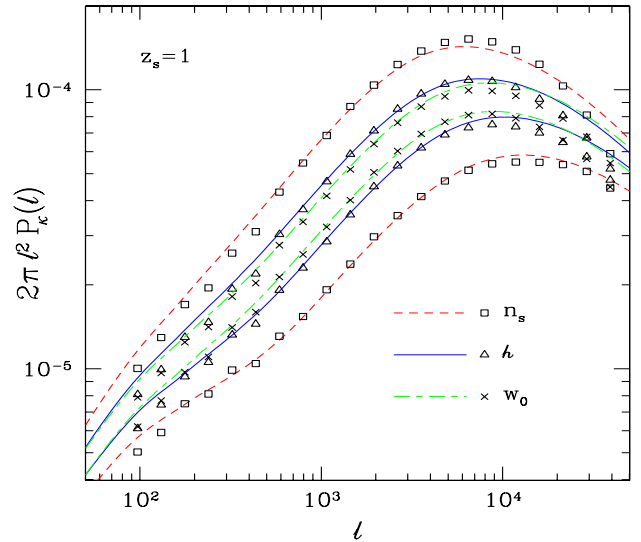}}
\end{center}
\caption{Convergence power spectrum for sources at redshift $z_s=1$, for
the six cosmologies given in Table~\ref{Table_cosmo}.
The points are the results from numerical simulations and the lines are the
predictions of our model.}
\label{fig_P_comp}
\end{figure}

\begin{figure}
\begin{center}
\epsfxsize=8.6 cm \epsfysize=7.5 cm {\epsfbox{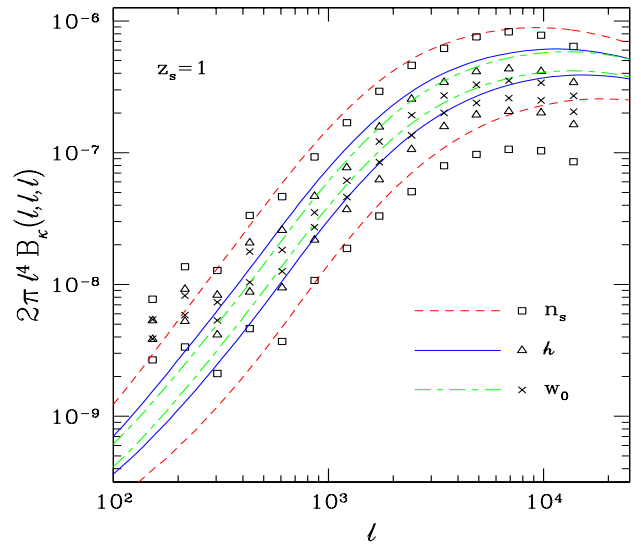}}
\end{center}
\caption{Convergence bispectrum for sources at redshift $z_s=1$, for
the six cosmologies given in Table~\ref{Table_cosmo}. 
The points are the results from numerical simulations and the lines are the
predictions of our model.}
\label{fig_B_comp}
\end{figure}

In this section we investigate the robustness of our model as we vary the
cosmological parameters. We consider six alternative cosmologies, 
where $n_s$, $\Omega_{\rm c}h^2$, and $w_0$ are modified by $\pm 10\%$
with respect to the fiducial cosmology used in the previous sections.
The values of the associated cosmological parameters are given in
Table~\ref{Table_cosmo} in App.~\ref{alternatives}.
We compare the predictions of our model with numerical simulations for these
six alternative cosmologies in Figs.~\ref{fig_P_comp} and \ref{fig_B_comp},
for the convergence power spectrum and bispectrum at $z_s=1$.
To avoid overcrowding the figures we do not plot the error bars of the numerical
simulations.
Each pair $n_s$, $\Omega_{\rm c}h^2$, and $w_0$ gives two curves that are roughly symmetric
around the fiducial cosmology result, since we consider small deviations
of $\pm 10\%$.
The deviations are largest for the $n_s$ case, which changes the
shape of the initial power spectrum as well as the normalization $\sigma_8$.
These six cases roughly cover the range that is allowed by current data
and the $n_s$ case is already somewhat beyond the observational bounds
\citep{Komatsu2011}. Thus, they provide a good check of the robustness of
our model for realistic scenarios.

We can see that the dependence of the convergence power spectrum on the 
cosmological parameters is well recovered by our model.
The agreement is somewhat less accurate for the bispectrum, since as for the
fiducial cosmology the theoretical predictions give more power on small scales
than is measured in the simulations. However, we recover the general trend
and obtain a reasonable match up to intermediate scales.
We can see that the simulation results for the convergence bispectrum
are scattered around the theoretical predictions on large scales because the
number of realizations is not large enough to converge. (Here we only use 40
realizations, whereas for the fiducial cosmology we used 1000 realizations.)
On small scales the effects of the finite resolution are larger than for the
power spectrum and the range of wavenumbers where simulation results are
reliable is clearly narrower in Fig.~\ref{fig_B_comp} than in Fig.~\ref{fig_P_comp}
(we clearly see unphysical deviations at low and high $\ell$).
This explains why the agreement appears worse for the bispectrum as this
is partly due to the larger numerical noise.

We obtain similar results for $z_s=0.6$ and $z_s=1.5$, as well as for other
cosmologies where we vary $A_s$ or $\Omega_{\rm de}$ by $\pm 10\%$.
This shows that our model and, more generally, models based on combinations
of perturbation theory and halo models provide a good modeling of the matter
distribution and of weak gravitational lensing effects and capture their dependence
on cosmology.
In particular,  Figs.~\ref{fig_P_comp} and \ref{fig_B_comp} show that
the accuracy of our model is sufficient to constrain $n_s$, $\Omega_ch^2$,
and $w_0$ to better than $10\%$.

\section{Conclusion}
\label{Conclusion}

In this article we have investigated the performance of current theoretical modeling
of the 3D matter density distribution with respect to weak-lensing statistics,
focusing on Fourier space statistics, specifically the convergence power
spectrum and bispectrum.
We find that for both quantities a model introduced in previous works
\citep{Valageas2011d,Valageas2011e} that combines 1-loop perturbation theory
with a halo model fares better than some other recipes based on fitting formulae
to numerical simulations or more phenomenological approaches.
It yields a reasonable agreement with numerical simulations and provides
a competitive approach, since it remains difficult and time-consuming to describe
a range of scales that spans three orders of magnitude or more by ray-tracing 
simulations.
Even though this particular model may still be improved, these results already
show that building on systematic and physically motivated models of the 3D matter 
distribution is a promising route to predict both 3D and 2D (i.e. projected) statistics.

An advantage of this approach, as compared with numerical simulations or global
fitting formulas, is that it provides a clear link between the observed weak-lensing
quantities, such as the convergence power spectrum or bispectrum on a given range
of multipoles $\ell$, and the various ingredients that govern the underlying 
3D matter density field. In particular, we can distinguish the scales associated with
the perturbative regime from those that probe nonperturbative features described
for instance by the 1-halo term. This is useful since the perturbative regime can be
predicted by rigorous and systematic approaches (e.g., in this paper we use
both standard perturbation theory and a peculiar resummation scheme, up to
one-loop order), which offers a well-controlled constraint on the cosmological
parameters. The nonperturbative regime relies on a more 
phenomenological approach, i.e. a halo model, so that the link to cosmological
parameters is somewhat weaker. Indeed, the accuracy is lower (because there
is no rigorous and systematic framework) and the model involves some
intermediate parameters, such as halo profiles, that show a weak dependence
on cosmology. Therefore, it can be useful to separate both regimes when deriving
constraints on cosmology from observations.
Moreover, the explicit link provided by such models allows one to use gravitational
lensing effects to probe the halo properties themselves.

We find that going to one-loop order brings a more significant improvement,
with respect to lowest-order perturbation theory, for the lensing bispectrum than
for the power spectrum. Moreover, while for the bispectrum the combination of
the 1-halo and 2-halo terms with the one-loop perturbative contribution (associated
with the 3-halo term) provides a reasonable match to simulations, for the power
spectrum there remains an intermediate regime that is not well described by the
simplest combination. This suggests that higher orders of the perturbative
expansion play a greater role for the power spectrum and should be taken into
account, or that the matching between the 2-halo and 1-halo contributions need be
improved. In our case, we used a simple geometric interpolation for the 3D
power spectrum and we find that it also provides good results for the weak
lensing power spectrum.
These features agree with previous results for the 3D density
power spectrum and bispectrum \citep{Sefusatti2010,Valageas2011e}.

As in 3D, the 2-halo term does not significantly contribute to the lensing bispectrum
for equilateral configurations. It only plays a role for very ``squeezed'' configurations
on weakly nonlinear scales, with a ratio of $\sim 100$ or more between the two
long sides and the smallest side of the Fourier-space triangle. This is also the regime
where we find a discrepancy between our predictions and the simulations.
Since both the 3-halo and the 1-halo terms provide good agreement with the 
simulations for all other regimes, where they dominate, this suggests that this 
discrepancy arises from a lower accuracy of the 2-halo term. This is not
really surprising since this contribution is a priori more intricate, because it
combines internal halo properties with larger-scale correlations.
Nevertheless, since this regime corresponds to extreme configurations it is
not a very serious practical problem.

Overall, we find that our model provides a good agreement with the numerical
simulations for a variety of cosmologies and gives a robust framework.
It could still be improved in various manners. First, the accuracy of the perturbative 
contribution may be increased by including higher orders beyond one-loop or
by using alternative resummation schemes. Second, the underlying halo model
could be refined to include substructures \citep{Sheth2003a,Giocoli2010},
deviations from spherical profiles \citep{Jing2002,Smith2006}, or the effect
of baryons \citep{Guillet2010}.
Next, the model could be generalized to non-Gaussian initial conditions, which should
yield distinctive signatures in the bispectrum \citep{Sefusatti2010}.
Another interesting generalization would be to include the effects of massive neutrinos.
Indeed, it is still difficult to model the damping associated with neutrino
free-streaming using numerical simulations and simple fitting formulae,
and analytic or semi-analytic prescriptions could be useful
\citep{Lesgourgues2009,Saito2009,Bird2012}.

In this first study we focused on Fourier-space statistics, the convergence power
spectrum and bispectrum, from which other statistics such as real-space
correlation functions can be derived. We shall present our results for such
real-space statistics in a companion paper \citep{Valageas2012}.

\begin{acknowledgements}

M.S. and T.N. are supported by a Grant-in-Aid for the Japan Society for Promotion of
Science (JSPS) fellows. This work is supported in part by the French
``Programme National de Cosmologie et Galaxies'' and the French-Japanese
``Programme Hubert Curien/Sakura, projet 25727TL'', the JSPS
Core-to-Core Program ``International Research Network for Dark Energy'', 
a Grant-in-Aid for Scientific Research on Priority Areas No. 467 ``Probing the Dark
Energy through an Extremely Wide and Deep Survey with Subaru
Telescope'', a Grant-in-Aid for Nagoya University Global COE
Program, ``Quest for Fundamental Principles in the Universe: from
Particles to the Solar System and the Cosmos'', and World Premier 
International Research Center Initiative (WPI Initiative), MEXT, Japan.
We acknowledge Kobayashi-Maskawa Institute for the Origin of
Particles and the Universe, Nagoya University for providing computing
resources.
Numerical calculations for the present work have been in part carried out 
under the ``Interdisciplinary Computational Science Program" in Center for 
Computational Sciences, University of Tsukuba, and also on Cray XT4 at 
Center for Computational Astrophysics, CfCA, of National Astronomical 
Observatory of Japan. 

\end{acknowledgements}

\appendix

\section{Six alternative cosmologies}
\label{alternatives}

\begin{table*}
\begin{center}
\begin{tabular}{c||c|c|c|c|c|c|c}
 & $\Omega_{\rm m}$ & $\Omega_{\rm de}$ & $\Omega_{\rm b}$ & $h$ & $n_s$
& $\sigma_8$ & $w_0$ \\ 
\hline\hline
fiducial & 0.238306 & 0.761694 & 0.0415995 & 0.732 & 0.958 & 0.759208 & -1 \rule[-0.23cm]{0cm}{0.6cm} \\ 
\hline\hline
$n_s+$ & - & - & - & - & 1.0538 & 0.920651 & - \rule[-0.23cm]{0cm}{0.6cm} \\ 
\hline
$n_s-$ & - & - & - & - & 0.8622 & 0.627354 & - \rule[-0.23cm]{0cm}{0.6cm} \\ 
\hline\hline
$\Omega_{\rm c}h^2+$ & - & - & 0.038429 & 0.7616 & - & 0.8149102 & - \rule[-0.23cm]{0cm}{0.6cm} \\ 
\hline
$\Omega_{\rm c}h^2-$ & - & - & 0.045347 & 0.7011 & - & 0.7014909 & - \rule[-0.23cm]{0cm}{0.6cm} \\ 
\hline\hline
$w_0+$ & - & - & - & - & - & 0.7813410 & -1.1 \rule[-0.23cm]{0cm}{0.6cm} \\ 
\hline
$w_0-$ & - & - & - & - & - & 0.7329135 & -0.9 \rule[-0.23cm]{0cm}{0.6cm} \\ 
\end{tabular}
\end{center}
\caption{The cosmological parameter values for fiducial and six
 alternative cosmologies. The parameters marked as ``-'' remain equal to
 the fiducial values.}
\label{Table_cosmo}
\end{table*}

We give in Table~\ref{Table_cosmo} the six alternative sets of cosmological
parameters that we consider in Sect.~\ref{Cosmology} to test the robustness
of our model.
The first line is our fiducial cosmology, used in other sections, while lines two
to seven give the cosmological parameters associated with the six scenarios
where $n_s$, $\Omega_{\rm c}h^2$, and $w_0$ are modified by $\pm
10\%$. The parameters marked as ``-'' remain equal to the fiducial values.
Several parameters in Table~\ref{Table_cosmo} may simultaneously vary
because we use the following set of independent parameters:
$A_s$, $n_s$, $\Omega_{\rm c}h^2$, $\Omega_{\rm b}h^2$, $\Omega_{\rm de}$,
$w_0$, and we assume the Universe is flat. 
For instance, if we change $\Omega_{\rm c}h^2$ keeping other parameters fixed,
$h$ has to be varied from the following equation: $h=\sqrt{\Omega_{\rm m}h^2/\Omega_{\rm
m}}=\sqrt{(\Omega_{\rm c}h^2+\Omega_{\rm b}h^2)/(1-\Omega_{\rm de})}$.
Then, $\Omega_{\rm c}$ and $\Omega_{\rm b}$ have to be varied as well,
because of $\Omega_{\rm b}=\Omega_{\rm b}h^2/h^2$ and the assumption of flat
Universe.
In the case of varying $n_s$ and $w_0$, $\sigma_8$ is changed because
other parameters (including $A_s$) are fixed.

\bibliographystyle{aa} % style aa.bst
\bibliography{ref3}

\end{document}